\documentclass[final,3p,11pt,pdflatex]{elsarticle}

\usepackage[utf8x]{inputenc}      % input font encoding
\usepackage[T1]{fontenc}          % output font encoding
\usepackage{lmodern}

\usepackage[obeyspaces]{url}
\usepackage{amsmath,amssymb,mathtools}
\usepackage{bm}
\usepackage{booktabs,tabularx}
\usepackage{graphicx}
\usepackage{xspace}
\usepackage[usenames]{xcolor}
\usepackage{listings}
\usepackage[absolute]{textpos}
\usepackage[many]{tcolorbox}
\usepackage{xparse}
\usepackage[font=small,labelfont=bf,format=plain,margin=0.05\textwidth]{caption}
\usepackage{bbm}
\usepackage{subcaption}
\usepackage{dsfont}
\usepackage[normalem]{ulem}

\renewcommand{\appendix}{%
    \setcounter{section}{0}
    \renewcommand*{\thesection}{\Alph{section}}
}

\renewcommand{\tfrac}{\frac}

\definecolor{green}{rgb}{0.13, 0.55, 0.13}
\definecolor{fscolor}{RGB}{44,118,255}

\usepackage[pdftitle={BubbleProfiler},
  pdfauthor={Peter Athron, Csaba Bal\'azs, Michael Bardsley,
    Andrew Fowlie, Dylan Harries, Graham White},
  pdfkeywords={BubbleProfiler, phase transitions, electroweak, baryogenesis},
  bookmarks=false, linktocpage, colorlinks=true,
  allcolors=fscolor]{hyperref}
\biboptions{sort&compress}
\bibstyle{elsarticle-num}

\allowdisplaybreaks

\newcommand{\bpversion}{1.0.0}

\newcommand{\bpgitrepo}{%
  \url{https://github.com/bubbleprofiler/bubbleprofiler}\xspace}
\newcommand{\bphomepage}{\bpgitrepo}
\newcommand{\bpdirname}{bubbleprofiler-\bpversion}
\newcommand{\bptarname}{v\bpversion}
\newcommand{\bptarpage}{%
  \url{https://github.com/bubbleprofiler/bubbleprofiler/archive/\bptarname.tar.gz}\xspace}
\newcommand{\docspage}{%
  \url{https://bubbleprofiler.github.io/}\xspace}

\newcommand{\bubblergitrepo}{%
  \url{https://github.com/bubbleprofiler/bubbler}}

\newcommand{\ab}{\texttt{AnyBubble}\@\xspace}
\newcommand{\boostminversion}{1.53.0\@\xspace}
\newcommand{\bp}{\texttt{BubbleProfiler}\@\xspace}
\newcommand{\cmake}{\texttt{CMake}\@\xspace}
\newcommand{\cmakeminversion}{2.8.12\@\xspace}

\newcommand{\code}[1]{\ifmmode\text{\nolinkurl{#1}}\else\nolinkurl{#1}\fi}

\newcommand{\cosmo}{\texttt{CosmoTransitions}\@\xspace}

\newcommand{\eigenminversion}{3.1.0\@\xspace}
\newcommand{\gev}{\ensuremath{\,\text{GeV}}}
\newcommand{\git}{\texttt{git}\@\xspace}
\newcommand{\ginacminversion}{1.6.2\@\xspace}
\newcommand{\gslminversion}{1.15\@\xspace}
\newcommand{\nloptminversion}{2.4.1\@\xspace}
\newcommand{\make}{\texttt{Make}\@\xspace}
\newcommand{\vevac}{\texttt{VEVacious}\@\xspace}

\newcommand{\figref}[1]{\figurename~\ref{#1}}
\newcommand{\subfigref}[1]{(\subref{#1})}
\newcommand{\secref}[1]{Section~\ref{#1}}
\newcommand{\appref}[1]{Appendix~\ref{#1}}
\newcommand{\tabref}[1]{\tablename~\ref{#1}}

\renewcommand{\refeq}[1]{Eq.~\ref{#1}}
\newcommand{\refeqs}[1]{Eq.s~\ref{#1}}
\newcommand{\refcite}[1]{Ref.~\cite{#1}}

\newcommand{\fv}{\ensuremath{\phi_f}}
\newcommand{\tv}{\ensuremath{\phi_t}}
\newcommand{\barrier}{\ensuremath{\phi_b}}
\newcommand{\rhomin}{\ensuremath{\rho_{\text{min}}}}
\newcommand{\rhomax}{\ensuremath{\rho_{\text{max}}}}
\DeclareMathOperator{\sech}{sech}
\DeclareMathOperator{\sinch}{sinch}

\NewDocumentEnvironment{OptionTable}{m m O{llXX}}{%
\table[tbh!]
  \tabularx{\textwidth}{#3}%
    \toprule
    Symbol & Default value & Allowed values & Description \\
    \midrule
}{\endtabularx\caption{#1}\label{#2}\endtable}

\makeatletter
\lstnewenvironment{numlstlisting}[1][]{%
  \lstset{%
    #1,
    numbers=left,
    firstnumber=auto,
    numberstyle=\tiny\sffamily}%
  \csname\@lst @SetFirstNumber\endcsname
}{%
  \csname \@lst @SaveFirstNumber\endcsname
}
\makeatother

\journal{Computer Physics Communications}
\begin{document}
\begin{frontmatter}
 \vspace*{0.5cm}
 \title{\Large\bf Bubbleprofiler: finding the field profile and action for cosmological  phase transitions}

\author[Monash]{Peter Athron}
\author[Monash]{Csaba Bal\'azs}
\author[Monash]{Michael Bardsley}
\author[Monash,Nanjing]{Andrew Fowlie}
\author[adelaide,prague]{Dylan Harries}
\author[Monash,Triumf]{Graham White}
\address[Monash]{ARC Centre of Excellence for Particle Physics at
  the Terascale, School of Physics, Monash University, Melbourne,
  Victoria 3800, Australia}
\address[Nanjing]{Department of Physics and Institute of Theoretical Physics, Nanjing Normal University, Nanjing, Jiangsu 210023, China}
\address[adelaide]{ARC Centre of Excellence for Particle Physics at
the Terascale, Department of Physics, The University of Adelaide,
Adelaide, South Australia 5005, Australia}
\address[prague]{Institute of Particle and Nuclear Physics, Faculty of
  Mathematics and Physics, Charles University in Prague, V
  Hole\v{s}ovi\v{c}k\'{a}ch 2, 180 00 Praha 8, Czech Republic}
\address[Triumf]{TRIUMF, 4004 Wesbrook Mall, Vancouver, British Columbia V6T 2A3, Canada}

  \begin{abstract}
    We present \bp, a C++ software package for finding field profiles in bubble walls and calculating the bounce action during phase transitions involving multiple scalar fields.
    Our code uses a recently proposed perturbative method for potentials with multiple fields and a shooting method for single field cases.
    \bp is constructed with modularity, flexibility and practicality in mind.
    These principles extend from the input of an arbitrary potential with multiple scalar fields in various forms, through the code structure, to the testing suite.
    \noindent After reviewing the physics context, we describe how the methods are implemented in \bp, provide an overview of the code structure and detail usage scenarios.
    We present a number of examples that serve as test cases of \bp and comparisons to existing public codes with similar functionality.
    We also show a physics application of \bp in the scalar singlet extension of the Standard Model of particle physics by calculating the action as a function of model parameters during the electroweak phase transition.
    \bp completes an important link in the toolchain for studying the properties of the thermal phase transition driving baryogenesis and properties of gravitational waves in models with multiple scalar fields.
    The code can be obtained from: \bphomepage.
  \end{abstract}

\begin{keyword}
phase transitions,
bounce solution,
Euclidean action, 
electroweak phase transition,
Higgs boson,
baryogenesis
% CsB:  https://www.elsevier.com/journals/computer-physics-communications/0010-46 "Ensure that the following items are present:" doesn't list PACs
% \PACS ???
\end{keyword}
\end{frontmatter}

% report numbers
\begin{textblock*}{10em}(\textwidth,1.5cm)
\raggedleft\noindent\footnotesize
CoEPP--MN--19--1
\end{textblock*}

\clearpage
\newgeometry{top=2.5cm,left=3cm,right=3cm,bottom=4cm,footskip=3em}
\section*{Program Summary}
\noindent
{\em Program title:} \bp\\[0.5em]
{\em Program obtainable from:} \bphomepage\\[0.5em]
{\em Distribution format:} tar.gz\\[0.5em]
{\em Programming language:} C++\\[0.5em]
{\em Computer:} Personal computer\\[0.5em]
{\em Operating system:} Tested on FreeBSD, Linux, Mac OS X \\[0.5em]
{\em External routines:} Boost library, Eigen library, GNU Scientific Library, NLopt library, GiNaC library \\[0.5em]
{\em Typical running time:} $< 1$ second for single field potentials, up to {\cal O}(10) seconds for potentials of several ($6+$) fields. \\[0.5em]
{\em Nature of problem:}
Find the field profile in the bubble wall (bounce solution) and Euclidean action for a cosmological phase transition by solving a set of coupled differential equations. \\[0.5em]
{\em Solution method:}
Direct shooting method for single field problems. Multiple field problems are solved using a perturbative algorithm which linearizes the bounce equations. \\[0.5em]
{\em Restrictions:} Currently unable to find bounce solutions for potentials of more than one fields where the vacua are nearly degenerate --- these are the so-called ``thin walled'' cases.

\clearpage
\tableofcontents

\newpage
\section{Introduction}
\label{sec:intro}
When a scalar field is in a local minimum, that is a deeper minimum of the potential exists separated from the local minimum by a barrier, the field will
eventually decay to the true vacuum through quantum tunneling
\cite{Coleman:1977py,Callan:1977pt,Linde:1980tt}.  Predicting the rate of such
a first order transition involves calculating the field profile for a critical
bubble.  This is a ubiquitous calculation in finite temperature quantum field
theory and, depending on the context, requires elements of particle physics and cosmology.  For
example, in electroweak baryogenesis \cite{Morrissey:2012db,White:2016nbo,
  Das:2000ft} the vacuum decays from an electroweak symmetric vacuum to a
broken one via bubble nucleation. Such a scenario requires new weak scale
physics to catalyze the electroweak phase transition (EWPT).  This has been
studied in detail for example in the minimal supersymmetric standard model
(MSSM) \cite{Lee:2004we,Balazs:2004ae,Carena:2008vj,Carena:2012np,
  Liebler:2015ddv,Huber:2001xf,Huber:2006ma}, the next-to-MSSM
\cite{Menon:2004wv,Balazs:2013cia,Akula:2017yfr,Davies:1996qn,Huber:2006wf,
  Kozaczuk:2014kva,Huber:2006ma,Bian:2017wfv}, as well as a variety of other
Beyond the Standard Model (BSM) scenarios including multistep transitions
\cite{Ramsey-Musolf:2017tgh,Inoue:2015pza,Patel:2013zla} and  effective field
theory (EFT) approaches \cite{Balazs:2016yvi,deVries:2017ncy,Grojean:2004xa,Kobakhidze:2015xlz,Huang:2015izx}.
The precise behavior of the bubble nucleation can determine the efficiency of
baryon production \cite{Balazs:2016yvi} as well as how quenched electroweak
sphalerons are which controls the degree to which the initial baryon yield is
washed out \cite{Patel:2011th}.

The recent discovery of gravitational waves \cite{Abbott:2016blz} has
also stirred interest in cosmic phase transitions, whether an
electroweak phase transition
\cite{Croon:2018new,Hashino:2018zsi,Chala:2018ari,Chen:2017cyc,
  Kang:2017mkl,Iso:2017uuu,Kobakhidze:2017mru,Beniwal:2017eik,Chao:2017vrq,Huang:2016odd,Huang:2017rzf,Chala:2018opy,Angelescu:2018dkk},
a dark sector phase transition \cite{Schwaller:2015tja,Shelton:2010ta,
  Chowdhury:2011ga,Croon:2018erz}, or phase transitions motivated by some other
ideas in particle physics or cosmology
\cite{Huber:2007vva,Balazs:2016tbi,
  Kamionkowski:1993fg,Apreda:2001us,Caprini:2007xq,Hindmarsh:2013xza,
  Caprini:2009fx,Espinosa:2010hh,Binetruy:2012ze,Huang:2017laj,Wan:2018udw,Mazumdar:2018dfl,Croon:2018kqn}.
In all cases the properties of the relic gravitational wave spectrum
are dependent on the precise details of these bubble wall profiles and
how they evolve
\cite{Huber:2008hg,Hindmarsh:2017gnf,Jinno:2017fby,Hindmarsh:2016lnk,
  Konstandin:2017sat,Weir:2017wfa}.

The stability of the Standard Model (SM) vacuum is also a related open problem.
The Higgs quartic self-coupling appears to turn negative at large scales
$\mathcal{O}(10^{10}\gev)$ when one studies its renormalization group (RG)
evolution to two loops \cite{Sher:1988mj,Camargo-Molina:2013sta,Blinov:2013fta,
  Swiezewska:2015paa,Bobrowski:2014dla,Hollik:2016dcm,MOLINA:2014uha}.  The
precise value is subject to experimental uncertainty in the top and Higgs masses
but a negative Higgs quartic implies that a catastrophic vacuum exists at very
large values of the Higgs field.  Whether this results in our vacuum being
unstable, metastable or stable on a cosmic time scale is an outstanding
theoretical and experimental problem \cite{Andreassen:2017rzq} which also
depends upon the maximum temperature in our cosmic history
\cite{Degrassi:2012ry,Rose:2015lna}.

Efficient publicly available codes exist for calculating the critical
temperature of a phase transition \cite{Basler:2018cwe}. However, the
accurate treatment of false vacuum decay is, unfortunately,
generically a numerically expensive problem.  Two publicly available
codes exist for calculating the decay of the false vacuum: \cosmo,
which is currently utilized by \vevac
\cite{Wainwright:2011kj,Camargo-Molina:2013qva}, and \ab
\cite{Masoumi:2016wot}.  \cosmo solves the bounce action using a path
deformation method, while \ab uses a multiple shooting method; we
discuss both of these methods in more detail in \secref{sec:tests}.
Various alternative methods for finding the bounce action have also
been proposed in the literature.  Since the action has a saddle point
at the bounce solution, which is a maximum with respect to
dilatations, \refcite{Claudson:1983et} extremizes dilatations of the
action appropriately to find the action at the bounce.  The authors of
Refs. \cite{Kusenko:1995jv, Kusenko:1996jn, Moreno:1998bq} and
\cite{John:1998ip} use optimization methods, by defining a
minimization function that describes departures from a modified
action.  Refs. \cite{Cline:1999wi, Dasgupta:1996qu} split the equation
of motion into two pieces in a similar manner to the path deformation
method of \refcite{Wainwright:2011kj}.  The authors of
Refs. \cite{Cline:1999wi} and \cite{Cline:1998rc} use a gradient
ascent/descent method to find the bounce.  In
\refcite{Konstandin:2006nd} the problem is solved on a lattice.
\refcite{Guada:2018jek} connects linear solutions in \refeq{eq:zero}
by approximating the potential by a polygon.
\refcite{Espinosa:2018hue} generalizes the single field, thin-wall
case by introducing a tunneling potential that connects smoothly the
false and true vacua and this method was very recently extended to the
multi-field case \cite{Espinosa:2018szu}.  The bounce action then is
expressed as a simple integral of the tunneling potential.  For the
single field case, machine learning techniques are used to find the
bounce in \refcite{Jinno:2018jov}.  Finally, and of greatest relevance
in the following, a new perturbative method was proposed in
\refcite{Akula:2016gpl}.

In this work we present our own easy to use
bounce solver \bp, which implements the algorithm in
\refcite{Akula:2016gpl} and a direct shooting method for single field
cases. We also compare and contrast \bp with \cosmo and \ab.  When
comparing the three codes we find that \bp is substantially faster
than \ab and faster than \cosmo for single field cases.  When
calculating the tunneling action the accuracy of \bp matches closely
that of \ab.

Going beyond the bounce solution to a fully-fledged calculation of the baryon
yield would require solving quantum Boltzmann equations in an inhomogeneous background, which is
highly non-trivial even in a toy model \cite{Cirigliano:2009yt,
  Cirigliano:2011di}. Approximate approaches focus on CP violation from a
semi-classical force \cite{Kainulainen:2001cn} or the CP violation in the
collision term can be estimated using the vev-insertion approach
\cite{Lee:2004we}. Some numerical and analytic methods have been proposed for
solving the transport equations in the latter case \cite{Chung:2009qs,
  White:2015bva}.

The structure of our paper is as follows. In \secref{sec:QuickStart}
we provide quickstart instructions for installing and running \bp.  We
describe the physical problem it solves in
\secref{sec:PhysicalProblem} and our approaches in the one-dimensional
and higher-dimensional cases in \secref{sec:shooting} and
\secref{sec:perturbative_algorithm}, respectively.  We present
detailed information about the structure of \bp in
\secref{sec:structure}.  We provide a detailed comparison of the
results from \bp with \cosmo, \ab and known analytic results in
\secref{sec:tests}, using our \code{bubbler} interface and scripts.
Finally in \secref{sec:ssm} we provide a quick physics application of
the code, looking at bubble nucleation in the scalar singlet model,
before concluding in \secref{sec:conclusions}.

\section{Quick start}
\label{sec:QuickStart}

\subsection{Requirements}

Building \bp requires the following:
\begin{itemize}
\item A C++11 compatible compiler (tested with \texttt{g++} 4.8.5 and higher,
  and \texttt{clang++} 3.3)
\item \cmake\footnote{See \url{http://cmake.org}.}, version \cmakeminversion
  or higher.
\item The GNU Scientific Library\footnote{See
  \url{http://www.gnu.org/software/gsl}.}, version \gslminversion or higher.
\item The \code{NLopt} library\footnote{See
  \url{https://nlopt.readthedocs.io/}.}, version \nloptminversion or
  higher.
\item GiNaC library\footnote{See \url{http://www.ginac.de}.},
  version \ginacminversion or higher.  Note that GiNaC also requires the CLN
  library\footnote{See \url{https://www.ginac.de/CLN/}.}, which
  may also need to be installed separately.
\item Eigen library\footnote{See \url{http://eigen.tuxfamily.org}.},
  version \eigenminversion or higher
\item Boost libraries\footnote{See \url{http://www.boost.org}.},
  version \boostminversion or higher, specifically:
  \begin{itemize}
  \item[*] \code{Boost.Program_options}
  \item[*] \code{Boost.Filesystem}
  \item[*] \code{Boost.System}
  \end{itemize}
\end{itemize}

\subsection{Downloading and running \bp}
The current release of \bp is available as a gzipped tarball from
\begin{center}
  \bphomepage ,
\end{center}
or alternatively the code may be obtained using the \git version
control system from the same location.  The documentation is hosted at
\begin{center}
  \docspage .
\end{center}
To download and uncompress \bp run at the command
line:
\begin{lstlisting}[language=bash]
$ wget (*\bptarpage*)
$ tar -xf (*\bptarname*).tar.gz
\end{lstlisting}
\bp uses the \cmake build system generator to configure the package and
generate an appropriate build system for the user's platform.  To build \bp on
a UNIX-like system with the \make build system installed as the default build
tool, run at the command line:
\begin{lstlisting}[language=bash]
$ cd (*\bpdirname*)
$ mkdir build
$ cd build
$ cmake ..
$ make
\end{lstlisting}
Note that performing an out-of-source build in a separate build directory as
illustrated above is recommended, but not compulsory.  The resulting library
and executable are located in the \code{lib/} and \code{bin/}
subdirectories of the main package directory, respectively.

Simple potentials can be run very quickly using the command line
interface \code{bin/run_cmd_line_potential.x} executable.  For example for the one-field potential,
\begin{equation}
V(x) = 0.1 \left((2 - x)^4 - 14 (2 - x)^2 + 24 (2 - x)\right),
\end{equation}
we can find the bounce action with the command,
\begin{lstlisting}[language={}]
bin/run_cmd_line_potential.x  --potential '0.1*((-x + 2)^4 - 14*(-x + 2)^2 + 24*(-x + 2))' --field 'x' --local-minimum 0.0 --global-minimum 5.0 --n-dims 3 
\end{lstlisting}
which results in the output
\begin{lstlisting}[language={}]
Potential: 0.1*((-x + 2)^4 - 14*(-x + 2)^2 + 24*(-x + 2))
Field: x
# Action: 54.112
\end{lstlisting}
More details about the command line interface are given in
\appref{sec:User-Options}.

\bp is also distributed with a suite of unit tests that may be used to
check that the compiled library behaves as expected by running the command:
\begin{lstlisting}[language=bash]
$ make check
\end{lstlisting}
Additionally, a number of small examples are provided with the downloaded
package.  These may be built by running:
\begin{lstlisting}[language=bash]
$ make examples
\end{lstlisting}
The resulting example programs may be found in the \code{bin/}
directory.

\section{The physics problem}
\label{sec:PhysicalProblem}

The Lagrangian of a single, real scalar field is
\begin{equation}
\mathcal{L} = \tfrac12 (\partial\phi)^2 - V(\phi).
\end{equation}
Switching to Euclidean time, $t \to -\imath \tau$, we find
\begin{equation}\label{eq:LE}
\mathcal{L} = -\tfrac12 (\nabla\phi)^2 - V(\phi),
\end{equation}
which leads to the equation of motion
\begin{equation}
\nabla^2 \phi = V^\prime(\phi).
\end{equation}
Assuming spherical symmetry the Laplace operator simplifies, resulting in
\begin{equation}\label{eq:one_dim_bounce}
\ddot \phi + \tfrac{n}{\rho}\dot\phi = V^\prime(\phi),
\end{equation}
where the dots and primes indicate derivatives with respect to $\rho$ and
$\phi$, respectively.  At zero temperature $n=3$ and
$\rho = \sqrt{\tau^2 + |x|^2}$, whereas at finite temperature $n=2$ and
$\rho = |x|$ \cite{Linde:1981zj}.
Following \refcite{Coleman:1977py}, we require that the field starts at rest,
\begin{equation}\label{eq:bcon1}
\dot\phi(\rho = 0) = 0,
\end{equation}
and ends at rest at the false vacuum,
\begin{equation}\label{eq:bcon2}
\phi(\rho \to \infty) = \fv \quad\text{and}\quad \dot\phi(\rho \to \infty) = 0.
\end{equation}
The trivial solution, $\phi(\rho) = \fv$, is physically irrelevant as it does not describe a phase transition.  We are interested in the so-called bounce action, found from the action corresponding to the Lagrangian in \refeq{eq:LE},
\begin{align}\label{eq:action}
S[\phi] &= \int dx^n d\tau \left(\tfrac12\nabla\phi^2 + V(\phi) - V(\fv) \right)\\
        &= \mathcal{S}_n \int_0^\infty d\rho \rho^n \left(\tfrac12\dot\phi^2 + V(\phi) - V(\fv) \right),
\end{align}
where the latter equality assumes a spherical symmetry and $\mathcal{S}_n$ is the surface area of an $n$-sphere, that is a sphere in $(n+1)$-dimensional space.  The bounce is in fact a saddle-point of the action \cite{Coleman:1987rm}.

This is equivalent to a classical mechanics problem for a point-like particle moving in an upturned potential $U(q) = -V(q)$ with an unusual friction term, $n\dot q/t$.  The equation of motion,
\begin{equation}
   \ddot q + \tfrac{n}{t}\dot q = U^\prime(q),
\end{equation}
may be found via the Euler-Lagrange method from the Lagrangian
\begin{equation}
    L(q, \dot q, t) = t^n \left(\tfrac12\dot q^2 - U(q)\right).
\end{equation}
We must delicately tune the initial position, $q(t=0)$, such that the particle rolls down a hill and balances exactly on top of another hill.  The unusual friction term falls with $1/t$.
Due to the mechanical analogy, in the following we refer to the argument of the field, $\rho$, as time.

For a quadratic potential of the form
\begin{equation}\label{eq:potential_form}
V(\phi) = V^{\phantom\prime}_0 + V^\prime_0 \phi + \tfrac{1}{2} m^2 \phi^2,
\end{equation}
the exact solution for $m^2 \neq 0$ is
\begin{equation}\label{eq:sol}
\phi(\rho) =  \frac{A}{(|m| \rho)^p} J_p\left(-\imath\sqrt{m^2} \rho\right) + \frac{B}{(|m| \rho)^p} Y_p \left(-\imath \sqrt{m^2} \rho\right) -\frac{V^\prime_0}{m^2},
\end{equation}
where $p = (n-1)/2$, $J_n$ is the order $n$ Bessel function of the first kind, and $Y_n$ is the order $n$ Bessel function of the second kind.  If, on the other hand, $m^2 = 0$, we find,
\begin{equation}
\phi(\rho) = A + \frac{B}{(n-1)} \frac{1}{\rho^{n - 1}}+ \frac{V^\prime_0}{2 (n + 1)} \rho^2 ,
\end{equation}
where the first and second terms are complementary solutions and the third term is the particular one.  If $m^2 \neq 0$, the qualitative behavior of the solution depends upon the sign of $m^2$.  For example, in the $n=2$ case, for $m^2 > 0$ we find
\begin{equation}\label{eq:gt}
\phi(\rho) = A\, \frac{\sinh\left(|m| \rho\right)}{|m|\rho} + B\, \frac{\cosh\left(|m| \rho\right)}{|m|\rho} - \frac{V^\prime_0}{m^2},
\end{equation}
where for simplicity we rescaled the constants $A$ and $B$. Whereas for $m^2 < 0$,
\begin{equation}\label{eq:lt}
\phi(\rho) = A\, \frac{\sin\left(|m| \rho\right)}{|m| \rho} + B\, \frac{\cos\left(|m| \rho\right)}{|m| \rho} - \frac{V^\prime_0}{m^2},
\end{equation}
where we again rescaled the constants $A$ and $B$. Finally for $m^2 = 0$,
\begin{equation}\label{eq:zero}
\phi(\rho) = A + \frac{B}{\rho} + \frac{V^\prime_0}{6} \rho^2.
\end{equation}
We will later utilize these solutions by Taylor expanding potentials to quadratic order to match the form in \refeq{eq:potential_form} and applying the boundary conditions in \refeq{eq:bcon1} and \refeq{eq:bcon2}.

The problem generalizes to $N$ scalar fields, in which case there are $N$
coupled second-order differential equations,
\begin{equation}\label{eq:n_dim_bounce}
\ddot \phi_i + \tfrac{n}{\rho}\dot\phi_i = \frac{\partial V}{\partial \phi_i},
\end{equation}
with $i=1,\ldots, N$.  The general $N$-field problem is significantly more challenging than the single-field case as the bounce solution may take a curved path through the field space.

\subsection{Thin- and thick-walled solutions}

Bounces are characterized by the time they spend close to the true vacuum.  This time-scale determines the magnitude of the action. There are two extremes: \emph{thin-walled} and \emph{thick-walled} bounces.  Thin-walled bounces occur when the true and false vacua are nearly degenerate,
\begin{equation}
\frac{V(\barrier) - V(\fv)}{V(\barrier) - V(\tv)} \simeq 1,
\end{equation}
where $\tv$ is the field value in the true vacuum and $\barrier$ is the value
of the field on the top of the barrier.  Note that this quantity is bounded by
zero and one.  In this extreme, losses to friction must be minimized to ensure
that there is sufficient energy to reach the false vacuum.  The bounce thus
sits close to the true vacuum until the time at which friction, which falls as
$1/\rho$, cannot stop it reaching the false vacuum. In the thin-walled limit,
the action diverges.

Thick-walled bounces, on the other hand, occur when the barrier between the
true and false vacua is negligible,
\begin{equation}
\frac{V(\barrier) - V(\fv)}{V(\barrier) - V(\tv)} \simeq 0.
\end{equation}
In this extreme, the initial potential energy must be similar to $V(\fv)$ to
ensure we do not overshoot the false vacuum.  The bounce spends limited time
sitting near the true vacuum. In \refeq{eq:measure} we introduce a measure of
the thickness/thinness of a bounce for the one-dimensional case.

\section{One-dimensional shooting}\label{sec:shooting}

\begin{figure}
    \centering
    \includegraphics[width=0.8\textwidth]{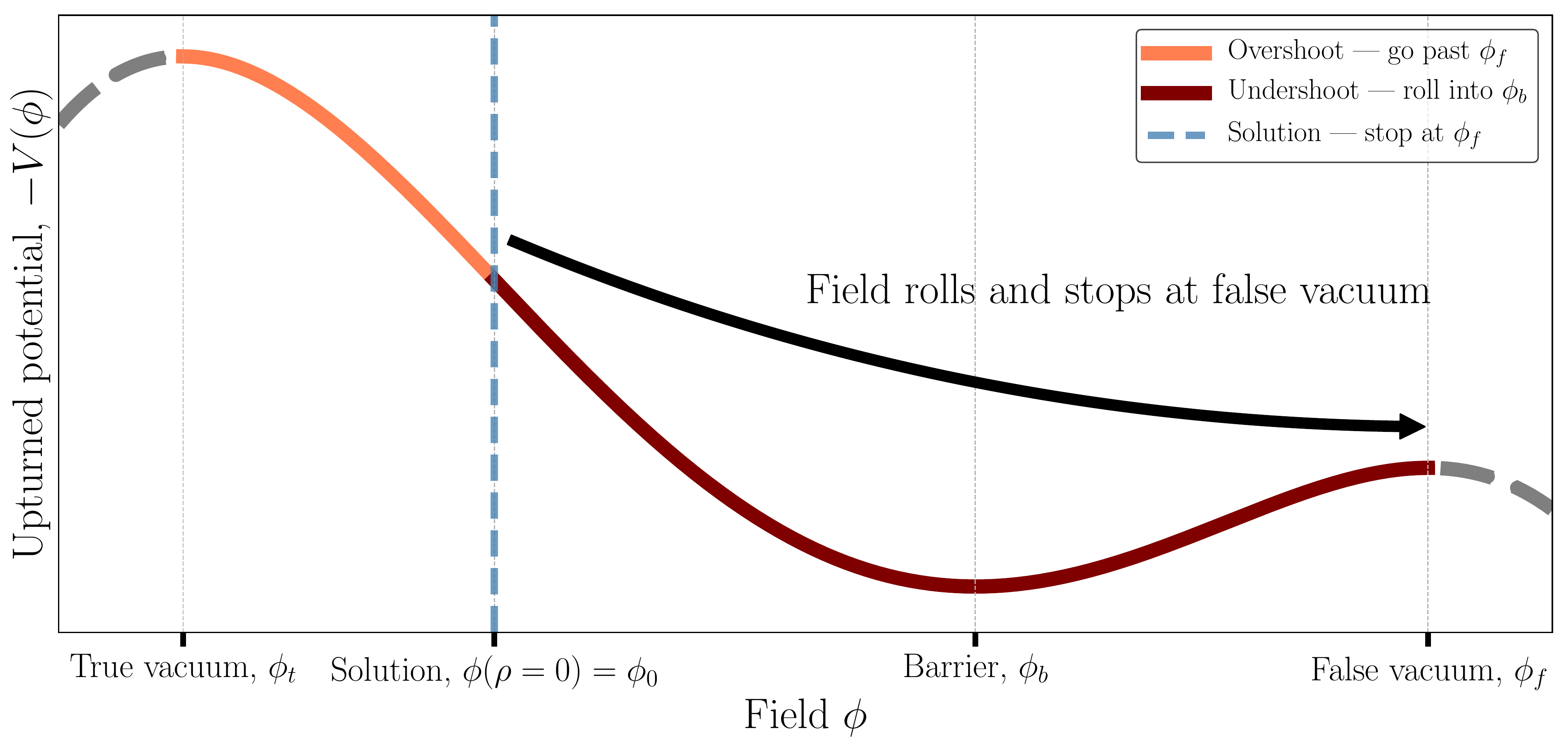}
    \caption{Illustration of shooting method for an upturned scalar potential
      with two minima. Beginning at rest in the orange region results in an
      overshot, whereas beginning in the brown region results in an undershot.
      We bisect between overshots and undershots until we find a solution
      (the blue dashed line). For the solution, the field rolls and comes to
      rest at the false vacuum.}
    \label{fig:shooting}
\end{figure}

To solve the one-dimensional bounce equation, that is the bounce equation for a single field, we use a shooting method, similar to that in \cosmo.  We guess an initial value for the field, $\phi_0 \equiv \phi(\rho = 0)$, evolve it in time, and check whether we \emph{overshot} or \emph{undershot}, that is whether we started too far up the maximum and rolled beyond the next maximum or started too close to the well lying between the maxima and rolled back into it.  This is illustrated in Fig.~\ref{fig:shooting}.  It was proven in \refcite{Coleman:1977py} that there is always at least one non-trivial solution.  The argument is roughly that a friction term is dissipative,
\begin{equation}
\frac{dE}{d\rho} = \frac{d}{d\rho} \left(\tfrac12\dot\phi^2 - V(\phi)\right) = -\tfrac{n}{\rho} \dot\phi^2 \le 0,
\end{equation}
thus we always undershoot if we begin with insufficient energy, e.g., close to the barrier.  On the other hand, it can be proven by inspecting \refeq{eq:sol} that if the field starts sufficiently close to the true vacuum, it remains close to the vacuum for an arbitrary time, after which the friction term may be neglected.  Thus, by energy conservation, we overshoot the true vacuum.

\subsection{Evolution with approximate solution}\label{sec:evolve}

For small $\rho$, the friction term dominates and the field evolves extremely slowly.  To avoid integrating the ODEs in this period, we first evolve the system by utilizing an approximate solution.  We expand the potential about our guess of $\phi_0$, neglecting terms higher than quadratic,
\begin{equation}
V(\phi - \phi_0) = V^{\phantom\prime}_0 + V^\prime_0 (\phi - \phi_0) + \tfrac12 m^2 (\phi - \phi_0)^2.
\end{equation}
This matches the form in \refeq{eq:potential_form} which is solved by
\refeq{eq:sol}.  Using the initial condition $\phi(\rho = 0) = \phi_0$ to fix
the integration constants, we obtain, e.g., in the $n=2$ case for $m^2 > 0$,
\begin{equation}\label{eq:approx_gt}
  \phi(\rho) - \phi_0 \approx \frac{V^\prime_0}{m^2} \left[\frac{\sinh(|m|\rho)}
    {|m|\rho} - 1\right],
\end{equation}
and for $n = 3$,
\begin{equation}\label{eq:approx_lt}
  \phi(\rho) - \phi_0 \approx \frac{V^\prime_0}{m^2} \left[\frac{2 I_1(|m|\rho)}
    {|m|\rho} - 1\right].
\end{equation}
The $m^2 = 0$ cases are simply the $m^2\to0$ limits of \refeq{eq:approx_gt} and
\refeq{eq:approx_lt}.  Note that using the initial condition
$\phi(\rho = 0) = \phi_0$, the constants of integration in \refeq{eq:zero}
must be zero.

We find the time, $\hat\rho$, at which the field has rolled a small fraction
$f$ towards the false vacuum,
\begin{equation}\label{eq:critical}
\left|\phi(\hat\rho) - \phi_0 \right| = f \left|\fv - \phi_0\right|,
\end{equation}
where $f$ is set by \code{Shooting::set_evolve_change_rel} and has a default
value of $10^{-2}$.  We solve this equation using approximate analytic
solutions.  E.g., in the $n = 2$ case we approximately invert hyperbolic sinc
using
\begin{equation}\label{eq:example_inverse_evolve}
\sinch^{-1}(x) \approx
\begin{cases}
	\sqrt{6 (x - 1)} & x < 1.5 \\
	-W_{-1}(-1 / (2 x)) & x \geq 1.5
 \end{cases} ,
\end{equation}
where $W_{-1}(x)$ is the negative branch of the Lambert-$W$ function.
Similarly, an approximate analytic solution is used in the $n = 3$
case\footnote{The implementation of these solutions may be found
  in the functions \code{double asinch(const double a)} and
  \code{double approx_root_eq_dim_4(const double a)} for the $n = 2$ and
  $n = 3$ cases, respectively.}.  Thus we solve for
$\hat\rho$, at which we know $\phi(\hat\rho)$ from \refeq{eq:critical}, and
calculate $\dot\phi(\hat\rho)$ from the analytic derivative of the approximate
solution.  There are two problematic cases, characterized by the thinness,
defined
\begin{equation}\label{eq:measure}
t \equiv \frac{f \left|\fv - \phi_0\right| m^2}{V^\prime_0}.
\end{equation}
If this quantity tends towards zero, we require a thick-wall solution and if it diverges, we require a thin-wall solution. Where necessary, we treat these cases with special asymptotic formulae.  In the case of thin-walled solutions, the asymptotic formulae are functions of the logarithm of the starting distance to the true vacuum (i.e., $\lambda$, as defined in \refeq{eq:thin_wall_reparam}).  This allows a numerical treatment of fine-tuned thin-wall cases. This is implemented in \code{Shooting::evolve}.

\subsection{Evolution with Runge-Kutta}

We take the field, $\phi(\hat\rho)$, and velocity, $\dot\phi(\hat\rho)$, and evolve them forwards in time with a controlled Runge-Kutta Dormand Prince method~\cite{DORMAND198019} implemented in \code{boost}.  This approach means that we do not solve ODEs in the period during which friction dominates and avoid the singularity in the ODE at $\rho = 0$.  This is implemented in \code{Shooting::ode}.

If the field is heading back towards the true vacuum or if there is insufficient kinetic energy,
\begin{equation}
\tfrac12\dot\phi^2 < -\left(V(\fv) - V(\phi)\right),
\end{equation}
we consider it an undershot and return \code{1}.  If the field has passed the false vacuum, we consider it an overshot and return \code{-1}.  This is implemented in \code{Shooting::shoot}.

We guess the initial step size in $\rho$ by guessing the characteristic size of the bubble,
\begin{equation} \label{eq:bubble_scale}
\Delta \rho = \frac{2\pi}{\sqrt{-V^{\prime\prime}(\phi_b)}}.
\end{equation}
This follows from approximating the potential at the barrier by a quadratic and finding the period of oscillations. This is implemented in \code{Shooting::bubble_scale}.  The initial step size is a fraction (\code{Shooting::set_drho_frac}) of this period, which is $10^{-3}$ by default.

\subsection{Bisection}

In \code{Shooting::shooting}, we bisect between overshots and undershots.  We bisect upon the variable
\begin{equation}\label{eq:thin_wall_reparam}
\lambda \equiv -\ln \frac{\phi_0 - \tv}{\barrier - \tv} ,
\end{equation}
between \code{Shooting::bisect_lambda_max = 5} and \code{0}, corresponding very near the true vacuum,
\begin{equation}
\frac{\phi_0 - \tv}{\barrier - \tv} = e^{-5} ,
\end{equation}
and the position of the barrier, $\barrier$.  If \code{Shooting::bisect_lambda_max = 5} is not an overshot, the range for the bisection is automatically shifted.  We stop once a relative precision of \code{Shooting::shoot_bisect_bits} significant bits is reached.  In thin-walled cases, the evolution in \secref{sec:evolve} depends only upon $\lambda$ and thus we may treat fine-tuned cases in which $\phi_0 - \tv$ is extremely small so long as the logarithm in \refeq{eq:thin_wall_reparam} is not so big that it cannot be represented by a \code{double}.

\subsection{Action}

Lastly, we calculate the action.  Because we find a stationary point of the action, we know that for $\phi(a\rho)$, the action must be extremized at $a=1$.  Thus we find that the contributions to the action from the kinetic term, $S_T$, and from the potential, $S_V$, are related (see e.g., \refcite{Masoumi:2016wot}) through
\begin{equation}
S_V = \frac{1 - n}{1 + n} S_T,
\end{equation}
where
\begin{align}
S_T &= \mathcal{S}_n \int_0^\infty d\rho \rho^n \, \tfrac12\dot\phi^2, \label{eq:kinetic_action}\\
S_V &= \mathcal{S}_n \int_0^\infty d\rho \rho^n \, \left[V(\phi) - V(\fv) \right].
\end{align}
Thus we may determine the action from $S_T$, $S_V$ or through a linear combination,
\begin{equation}
S = S_T + S_V = \frac{2}{1 + n} S_T = \frac{2}{1 - n} S_V.
\end{equation}
We find that calculating the action from summing only the kinetic term is more accurate,
especially for thick-walled solutions, and avoids evaluations of the potential,
which may be computationally expensive.  For the interior of the bubble wall,
at $\rho \le \hat\rho$, we approximate the action using analytic integration
of the approximate solution in \refeq{eq:sol}.  We treat thin-walled cases by
a special asymptotic formula.

For the bubble itself, we evolve the fields one more time from $\phi(\hat\rho)$,
this time summing the action using a trapezoid rule, with an increment
$\Delta\rho$ determined by the adaptive Runge-Kutta method.  We stop the
evolution once we undershoot, overshoot, or arrive at the false vacuum to within
a relative tolerance of \code{Shooting::action_arrived_rel}.  This is implemented
in \code{Shooting::action}.

The accuracy difference between calculating the kinetic and potential terms stems from the fact that summing the potential term involves
a cancellation between positive and negative contributions.  When the field
is on top of the false vacuum, the integrand is negative.  When it is in the
well, it is positive.  Thus, especially for thick-walled solutions that slowly roll through the well, we require a precise subtraction
between the positive and negative contributions, which can be numerically challenging. The integrand of the
kinetic term, on the other hand, is always positive.

\section{Perturbative method for multidimensional potentials}
\label{sec:perturbative_algorithm}

When the potential is a function of more than one field, direct shooting is
no longer possible as the enlarged space of initial conditions precludes an
approach based on bisection.  We instead apply the Newton-Kantorovich method
\cite{NKtheorem}, as discussed in \refcite{Akula:2016gpl}.  In this approach,
the original nonlinear problem is solved using an iterative algorithm in which
successive corrections are applied to a judiciously chosen initial guess for
the bounce solution.  At each step of the iteration, the corrections are
determined by solving a system resulting from linearizing the equations of
motion, and hence are easily calculated using standard numerical methods.
The iteration continues until the estimated action and values of the fields
converge to within the desired tolerance.  Unlike the shooting method in
\secref{sec:shooting}, this method is applicable to both the single and
multi-field cases.

\subsection{Ansatz}\label{sec:ansatz}

The first step in the perturbative algorithm is to construct an initial guess
or ansatz for the solution.  To construct this ansatz, we typically assume that
the path is a straight line between the true and false vacua.  On the straight
line path, the potential is a one-dimensional function of the path length.  We
provide two options for the profile along that path, which is our ansatz, and
also allow for arbitrary ansatzes:
\begin{description}

\item[Shooting] The most straightforward way to construct an ansatz is to solve
  this reduced, one-field problem using the shooting method of
  \secref{sec:shooting}.

\item[Look-up tables] A less computationally expensive option is described in
  \refcite{Akula:2016gpl}, which proposes an analytic ansatz based on the
  idea that the potential between the true and false vacua can be approximated
  by a fourth-order polynomial.  Once that is done, through reparameterizations
  the potential can be recast in terms of a single parameter, $\alpha$ (see
  \secref{sec:reparam}).

The so-called kink solutions
\begin{equation}\label{eq:kink_ansatz}
    \phi(\rho) \approx \phi_0
\left[
     1 - \tanh \left(\frac{\rho - \delta(\alpha)}{w(\alpha)} \right)
     - \frac{L}{w(\alpha)} \sech^2\left(\frac{\delta(\alpha)}{w(\alpha)}\right)
     e^{-\rho / L}
\right],
\end{equation}
are approximate solutions to the tunneling problem for a fourth-degree
polynomial potential of the form given in \refeq{eq:1d_E_alpha}.  Here
the functions $\delta(\alpha)$ and $w(\alpha)$
characterize respectively the location and width of the bubble wall.  They are
found by interpolation from look-up tables built from numerical fits based on
solutions using the methods of \secref{sec:shooting}.  The resulting ansatz
typically approximates the numerical bounce solution for the potential
in \refeq{eq:1d_E_alpha} with an absolute discrepancy of at most $0.003$.
We fix $L = 1 / (\fv - \tv)$. Note that $\phi(\rho = 0)
\approx \phi_0$ and that we have added a term to the usual kink solution to
ensure that $\dot\phi(\rho = 0) = 0$.

\item[Text file] An arbitrary ansatz may also be provided in a text file.  This
  may be desirable if a straight-line between the true and false vacua is a poor
  ansatz, regardless of the behavior along that path with respect to $\rho$.
  See Appendix \ref{sec:ansatz_cli_options} for a description of the required
  file format.

\end{description}

\subsection{Perturbative corrections} \label{sec:perturbative_corrections}

Starting from an initial ansatz constructed in the manner described above,
we then proceed to compute a series of approximations $\phi_i^{(j)}(\rho)$ to
the exact solution $\phi_i(\rho)$ for the $i^{\text{th}}$ field, $i = 1, \ldots,
N$.  The $j^{\text{th}}$ iterate is given by
\begin{equation} \label{eq:nk-iterates}
  \phi_i^{(j)}(\rho) = \phi_i^{(j-1)}(\rho) + \epsilon_i^{(j-1)}(\rho) ,
  \quad i = 1, \ldots, N, \, j = 1, 2, \ldots
\end{equation}
where $\epsilon_i^{(j)}(\rho)$ denotes a correction to the previous iterate
and $\phi_i^{(0)}(\rho) \equiv A_i(\rho)$.  Provided that our initial guess is
carefully chosen, the corrections $\epsilon_i^{(j)}(\rho)$ are expected to
satisfy $|\epsilon_i^{(j)}(\rho)| \ll |\phi_i^{(j)}(\rho)|$ for each field $i$
and for all $\rho$.

The corrections $\epsilon_i^{(j)}(\rho)$ at each step of the iteration are
determined by requiring that $\phi_i^{(j+1)}(\rho)$ should approximately satisfy
the classical equations of motion.  Substituting \refeq{eq:nk-iterates} into
\refeq{eq:n_dim_bounce} and Taylor expanding the scalar potential yields
\begin{equation} \label{eq:epsilon_eom}
  \frac{\partial^2 \epsilon_i^{(j)}}{\partial \rho^2}
  + \frac{n}{\rho} \frac{\partial \epsilon_i^{(j)}}{\partial \rho}
  - \sum_{k} \left . \frac{\partial^2 V(\phi)}{\partial \phi_i \partial \phi_k}
  \right |_{\phi^{(j)}} \epsilon_k^{(j)} = B_i^{(j)}(\rho) + O(\epsilon^2) ,
\end{equation}
where the inhomogeneous terms $B_i^{(j)}(\rho)$ are given by
\begin{equation} \label{eq:epsilon_inh}
  B_i^{(j)}(\rho) = \left . \frac{\partial V(\phi)}{\partial \phi_i}
  \right |_{\phi^{(j)}} - \frac{\partial^2 \phi_i^{(j)}}{\partial \rho^2}
  - \frac{n}{\rho} \frac{\partial \phi_i^{(j)}}{\partial \rho} .
\end{equation}
Upon neglecting those terms that are $O(\epsilon^2)$ or higher, we arrive
at a linear system of differential equations for the corrections
$\epsilon_i^{(j)}$.  Note that this is similar to the approach in
\refcite{Konstandin:2006nd}, in which one Taylor expands to first order
about an ansatz found by solving the bounce equation without friction.  The
necessary boundary conditions for the corrections are obtained by substituting
the definition of $\epsilon_i^{(j)}$ into the boundary conditions for the fields
$\phi_i$,
\begin{equation} \label{eq:perturbation_bcs}
  \epsilon_i^{(j)}(\rho \to \infty) = {\fv}_i - \phi_i^{(j)}(\rho \to \infty) ,
  \quad
  \dot{\epsilon}_i^{(j)}(\rho = 0) = -\dot{\phi}_i^{(j)}(\rho = 0) .
\end{equation}
In doing this we have gained the advantage of replacing $N$ nonlinear coupled
equations with $N$ linear coupled equations.  The resulting linear boundary
value problem may then be efficiently solved using standard numerical
techniques.  In particular the $N$-dimensional generalization of the shooting
method described in section 18.1 of
\refcite{VetterlingNumericalRecipesExample2017} is directly applicable.  We
review this technique in \secref{sec:multiple_shooting}.

\begin{figure}[h]
    \centering
    \begin{subfigure}[b]{0.48\textwidth}
        \centering
        \includegraphics[width=\textwidth]{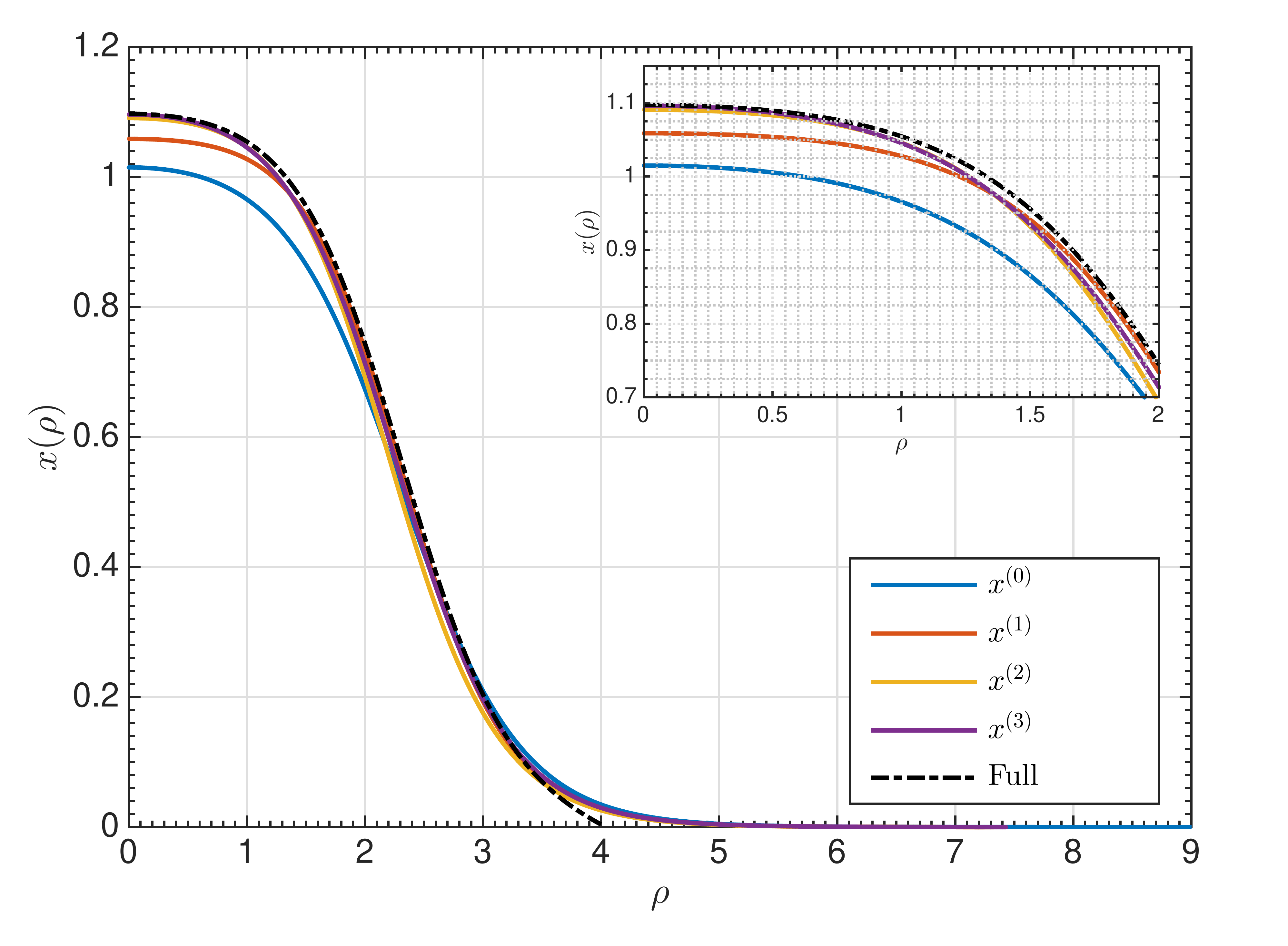}
    \end{subfigure}
    \hfill
    \begin{subfigure}[b]{0.48\textwidth}
        \centering
        \includegraphics[width=\textwidth]{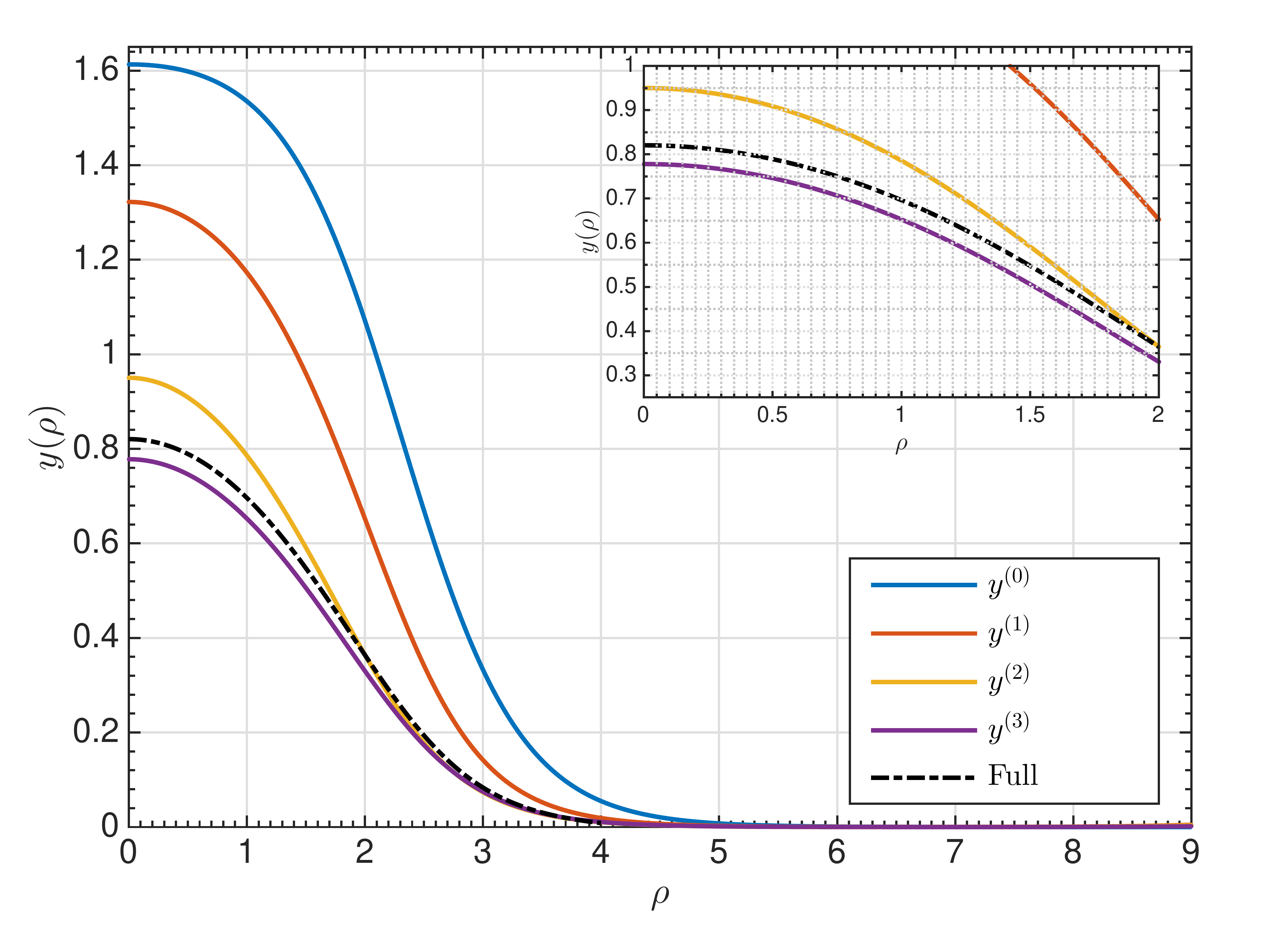}
    \end{subfigure}
    \vskip\baselineskip
    \begin{subfigure}[b]{0.48\textwidth}
        \centering
        \includegraphics[width=\textwidth]{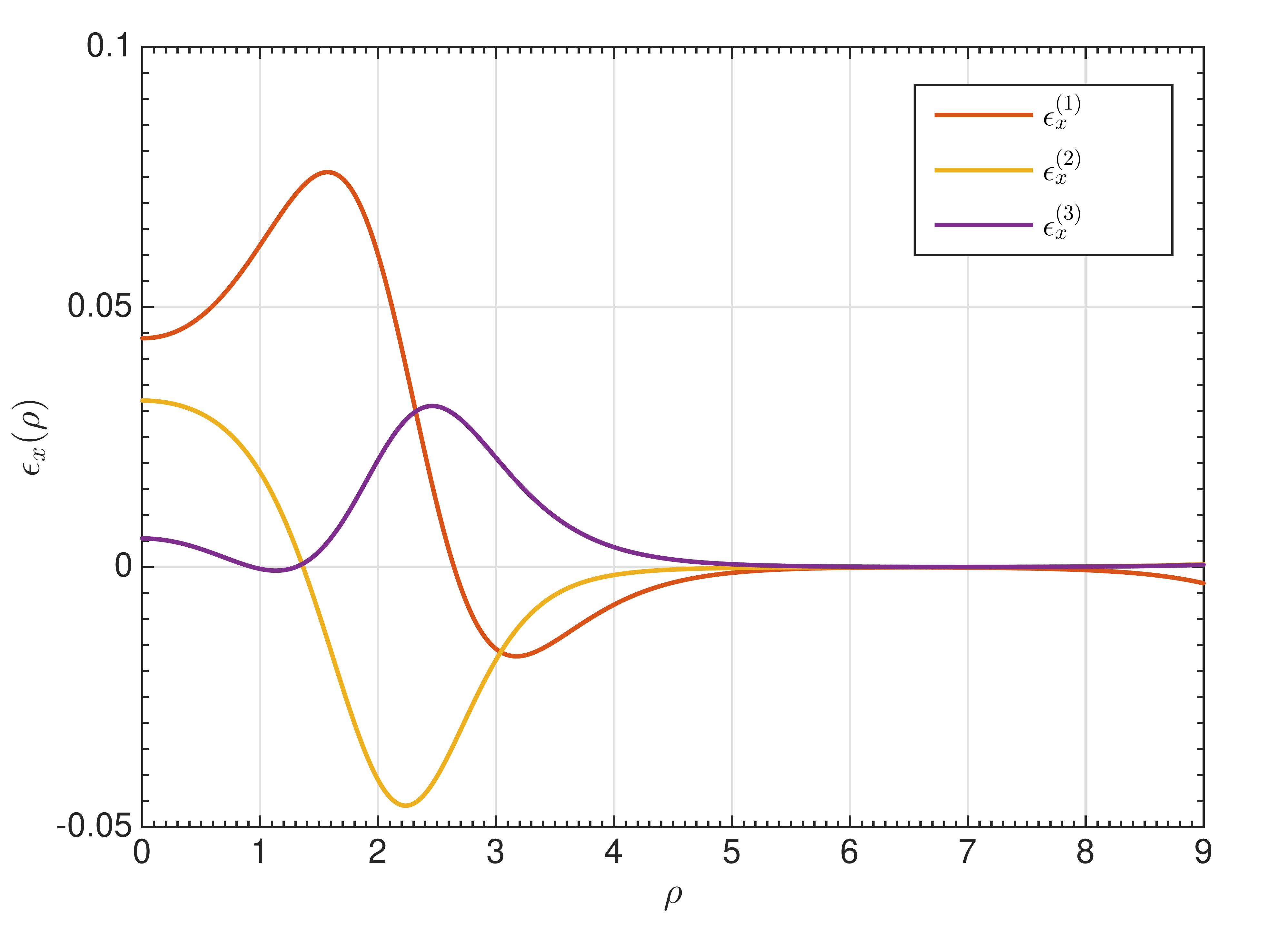}
    \end{subfigure}
    \quad
    \begin{subfigure}[b]{0.48\textwidth}
        \centering
        \includegraphics[width=\textwidth]{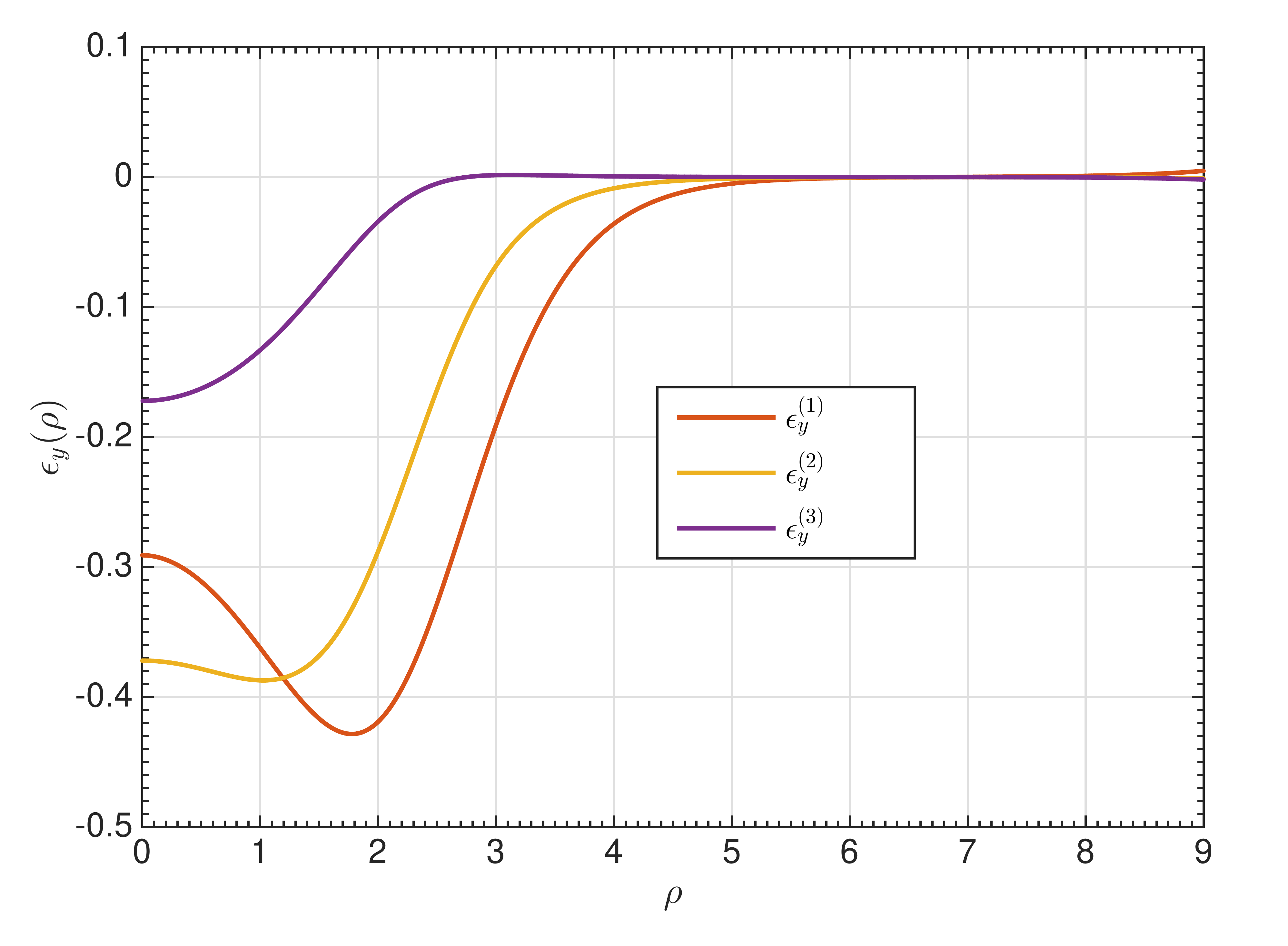}
    \end{subfigure}
    \caption[]{%
      The perturbative algorithm for a two field potential, $V(x, y)$. The top
      two panels show the field profiles $x(\rho)$ and $y(\rho)$ at each
      iteration starting from a kink ansatz (solid blue), and converging to
      the true solution (dotted black). The bottom two panels show the
      corrections, $\epsilon$, computed at each step by solving
      \refeq{eq:epsilon_eom}. Figure reproduced from \cite{Akula:2016gpl}.}
    \label{fig:perturbative_method}
\end{figure}

In general, the convergence of this iterative procedure to the true solution
depends on the quality of our initial guess for the solution.  In particular,
the guess for the solution must be sufficiently accurate so that neglecting
the $O(\epsilon^2)$ terms in \refeq{eq:epsilon_eom} is justified.
A heuristic condition for the validity of this approach is that the missing
terms are smaller in magnitude than the terms kept in the Taylor series.
However, since we do not directly control the size of $\epsilon$, violations
of this rule are somewhat inevitable for some values of $\epsilon(\rho)$.
The violations, though, do not necessarily prevent our algorithm from reaching
a bounce solution \cite{Akula:2016gpl}.  When this is not the case and
a poor initial guess is responsible for a failure of the algorithm, one may
of course attempt to remedy the situation by employing an alternative ansatz.
Provided that the chosen ansatz satisfies a set of
sufficient conditions that depend only on the particular system at hand
and the initial guess itself, then the iteration is guaranteed to converge
and, for a sufficiently accurate initial guess, is expected
to do so with a quadratic rate of convergence \cite{NKtheorem}.
We illustrate this for a typical two-field example in
\figref{fig:perturbative_method}.  The corrections $\epsilon$ reduce with each
iteration, quickly converging to a bounce solution.

In practice, we terminate the iteration once the iterates satisfy a set of
convergence criteria to within a predetermined numerical tolerance.
There are several possible convergence criteria.  \refcite{Akula:2016gpl}
suggests the vanishing of the inhomogeneous terms, \refeq{eq:epsilon_inh}, to
within a specified tolerance.  These, of course, vanish for the exact bounce
solution.  By default, we instead specify thresholds for the relative changes
in the action, $\kappa_S$, and initial values of the fields,
$\kappa_\phi$,\footnote{In \bp, $\kappa_S$ and $\kappa_\phi$ are configured
  respectively by the \code{rtol_action} and \code{rtol_fields} parameters.}
and store the Euclidean action $S^{(j)}$ and starting position of each field
$\phi^{(j)}_i(0)$ at each iteration. When both of the following conditions are
met:
\begin{equation} \label{eq:convergence_action}
  \frac{\left|S^{(j)} - S^{(j-1)}\right|}{\max\left(S^{(j)},S^{(j-1)}\right)}
  < \kappa_S,
\end{equation}
\begin{equation} \label{eq:convergence_fields}
  \max_i \left(\frac{\left|\phi^{(j)}_i(0) - \phi^{(j-1)}_i(0)\right|}{\max
    \left(\phi^{(j)}_i(0), \phi^{(j-1)}_i(0)\right)}\right) < \kappa_\phi
\end{equation}
the algorithm terminates and returns the Euclidean action and field profiles.

\subsection{Multiple shooting method}\label{sec:multiple_shooting}

As described above, at each step of the perturbative algorithm, one needs to
compute correction functions $\epsilon_i(\rho)$ by solving the boundary value
problem (BVP) in \refeq{eq:epsilon_eom} and \refeq{eq:perturbation_bcs} for
a system of $N$ linear differential equations (once terms that are
$O(\epsilon^2)$ or higher are neglected).  This section will outline a
straightforward approach to solving these equations that generalizes the
shooting method of \secref{sec:shooting} to multiple fields.  However, since
we are now operating on linear equations, the algorithm converges in a single
step.  To simplify the notation, in the following we describe the shooting
method as applied to determine the corrections $\epsilon_i(\rho)$ to the
initial ansatz $A_i(\rho)$; the equivalent expressions for later stages of
the iteration follow by replacing $A_i(\rho) \rightarrow \phi_i^{(j)}(\rho)$
and $\epsilon_i(\rho) \rightarrow \epsilon_i^{(j)}(\rho)$.

The boundary conditions fix $\dot{\epsilon}_i(0) = -\dot{A}_i(0)$.  Solving the
BVP means finding initial values $\epsilon_i(0) \equiv \epsilon_{0i}$ such that
integrating from the initial conditions yields a solution satisfying
\begin{equation}
\epsilon_i(\rho\to\infty) = {\fv}_i - A_i(\rho\to\infty).
\end{equation}

In fact, we work on a finite domain $[\rhomin, \rhomax]$ to avoid problems due
to the $n/\rho$ term in \refeq{eq:n_dim_bounce}.  Since we use
\refeq{eq:kinetic_action} to calculate the action, and our boundary conditions
require that the derivatives $\dot{\phi}_i(\rho)$ vanish as $\rho \rightarrow
\infty$ and as $\rho \rightarrow 0$, for sufficiently small $\rhomin$ and large
$\rhomax$ the action calculation remains accurate.  On the finite domain, the
boundary conditions become
\begin{equation}\label{eq:perturbation_bcs_finite}
  \epsilon_i(\rhomax) = {\fv}_i - A_i(\rhomax), \quad
  \dot{\epsilon}_i(\rhomin) = -\dot{A}_i(\rhomin).
\end{equation}

We want to solve the second order system
\begin{equation}
  \ddot{\epsilon}_i + \frac{n}{\rho}\dot{\epsilon}_i
  - \left.\sum_j\frac{\partial^2 V(\phi)}{\partial \phi_i \partial \phi_j}
  \right|_{A(\rho)} \epsilon_j = B_i(\rho),
\end{equation}
with $B_i(\rho)$ as given in \refeq{eq:epsilon_inh}.
This may be reduced in the usual manner to a system of $2N$ first order
equations for $\epsilon_i$ and the new variables
%
%\begin{align}
\begin{equation}
\zeta_i \equiv \dot{\epsilon_i} ,
\end{equation}
implying
\begin{equation}
  \dot{\zeta}_i = B_i(\rho) + \left.\sum_j \frac{\partial^2 V(\phi)}
      {\partial \phi_i \partial \phi_j}\right|_{A(\rho)} \epsilon_j
      - \frac{n}{\rho}\zeta_i.
\end{equation}
%\end{align}
We solve this system using a form of Newton's method
\cite{VetterlingNumericalRecipesExample2017}, suitably generalized
for more than one field.

It is instructive to write the system in matrix form,
\begin{equation}
\label{eq:perturbations_matrix}
\dot{\bm{x}}(\rho) = C(\rho) \, \bm{x}(\rho)
+ \bm{b}(\rho),
\end{equation}
where $\bm{x}(\rho) = (\epsilon_1,...,\epsilon_N,\zeta_1,...,\zeta_N)^T$,
$\bm{b}(\rho) = (0,\ldots,0,B_1(\rho),\ldots,B_N(\rho))^T$, and
$C(\rho)$ is a $2N\times 2N$ block matrix:
\begin{equation}
C(\rho) = \begin{pmatrix}
0 & \mathds{1} \\
J & P
\end{pmatrix},\ P_{ij} = -\frac{n}{\rho} \delta_{ij},\ J_{ij} =
\frac{\partial^2 V(\phi)}{\partial \phi_i \partial \phi_j}\Big|_{A(\rho)}.
\end{equation}
In the above, $\mathds{1}$ is the $N\times N$ identity matrix and $\delta_{ij}$
is the Kronecker delta.  The general solution of \refeq{eq:perturbations_matrix}
is given by the Peano-Baker series \cite{BaakePeanoBakerseries2011},
\begin{equation}
\label{eq:peano_baker_system}
\bm{x}(\rho) = S[\rho,\rhomin]\,\bm{x}(\rhomin)
+ \int_{\rhomin}^{\rho}S[\rho,s]\bm{b}(s) ds,
\end{equation}
where $S[\rho, \rho^{\prime}]$ is a linear operator for all
$\rho > \rho^{\prime}$.  The important thing about this
representation is that if we write our initial conditions as
\begin{equation}
\bm{x}(\rhomin) = \begin{pmatrix}
\bm{\epsilon}(\rhomin) \\
\bm{\zeta}(\rhomin)
\end{pmatrix},
\end{equation}
and note that the components of $\bm{\zeta}(\rhomin)$ are fixed by the
second set of boundary conditions in \refeq{eq:perturbation_bcs_finite}, we can
consider integrating the equations from $\rhomin$ to $\rhomax$ as an affine map
$\mathcal{E}$ between the initial and final values for $\bm{\epsilon}$
only,
\begin{equation}
  \mathcal{E}\left[\bm{\epsilon}(\rhomin)\right] \equiv \bm{\epsilon}(\rhomax)
  = \tilde{S}[\rhomax,\rhomin]\,\bm{\epsilon}(\rhomin) + B[\rhomin, \rhomax].
\end{equation}
Here $B[\rhomin,\rhomax]$ is the first $N$ entries of the inhomogeneous term in
\refeq{eq:peano_baker_system}, and $\tilde{S}[\rhomax,\rhomin] \equiv \pi \circ
S[\rhomax,\rhomin]$ where $\pi$ denotes projection onto the first $N$
coordinates.  We can obtain an estimate for $\tilde{S}[\rhomax,\rhomin]$ with
$N + 1$ integrations.  Fixing an initial guess $\bm{\epsilon}^0$, we may
write
\begin{equation}
  \tilde{S}[\rhomax,\rhomin]_{ij} =
  \frac{\partial \mathcal{E}_i}{\partial \epsilon_j}
  \approx \frac{\mathcal{E}_i(\epsilon^0_1,...\epsilon^0_j
    + \Delta \epsilon,...,\epsilon_N)
    - \mathcal{E}_i(\epsilon^0_1,...,\epsilon_N)}{\Delta \epsilon}.
\end{equation}

To satisfy the boundary condition $\bm{\epsilon}(\rhomax) = \bm{\fv}
- \bm{A}(\rhomax) \equiv \hat{\bm{\epsilon}}$, we need to find a
correction $\delta\bm{\epsilon}$ to our initial guess such that
$\mathcal{E}(\bm{\epsilon}^0
+ \delta \bm{\epsilon}) = \hat{\bm{\epsilon}}$.  Since
\begin{align}
  \mathcal{E}(\bm{\epsilon}^0 + \delta\bm{\epsilon})
  - \mathcal{E}(\bm{\epsilon}^0) &=
  \tilde{S}[\rhomax,\rhomin]\delta\bm{\epsilon} \\
&= \hat{\bm{\epsilon}} - \mathcal{E}(\bm{\epsilon}^0), \nonumber
\end{align}
$\delta\bm{\epsilon}$ can be obtained by solving the linear system
$\tilde{S}[\rhomax,\rhomin]\delta\bm{\epsilon} = \hat{\bm{\epsilon}}
- \mathcal{E}(\bm{\epsilon}^0)$.  The algorithm to compute the next
correction $\bm{x}(\rho)$ is then complete after a final integration
from the corrected initial conditions,
\begin{equation}
\bm{x}(\rhomin) = \begin{pmatrix}
\bm{\epsilon}^0 + \delta \bm{\epsilon} \\
-\dot{\bm{A}}(\rhomin)
\end{pmatrix}.
\end{equation}

\section{\bp structure}\label{sec:structure}

\bp is a C++ software package for finding the bounce solution and
outputting the bounce action and field profiles. \bp is designed so
that multiple methods for finding the bounce solution may be
implemented.  Currently the main bounce solver is a perturbative
algorithm described in \secref{sec:perturbative_algorithm}, which uses
the multiple shooting method described in
\secref{sec:multiple_shooting} to solve the linearized correction
equations.  Additionally, a fast implementation of the nonlinear
direct shooting method outlined in \secref{sec:shooting} is available
to solve single-field problems.  We intend \bp to become a mature
software package suitable for widespread use in scientific research,
and integration into larger phenomenology frameworks.  Reflecting this
ambition, \bp includes:
\begin{itemize}
\item A comprehensive suite of unit tests.
\item Detailed Application Programming Interface (API) documentation.
\item A modular architecture designed using the dependency injection principle
  for maximum flexibility.
\item A large selection of example scripts illustrating the use of the code.
\item A continuous integration pipeline which automatically builds, tests, and
  updates the API documentation when changes are made to the code.
\end{itemize}
Detailed and up to date API documentation, including instructions for the
installation and usage of the code are available at \docspage.

\subsection{\bp Architecture}  

\begin{figure}[h]
\centering
\includegraphics[width=0.9\textwidth]{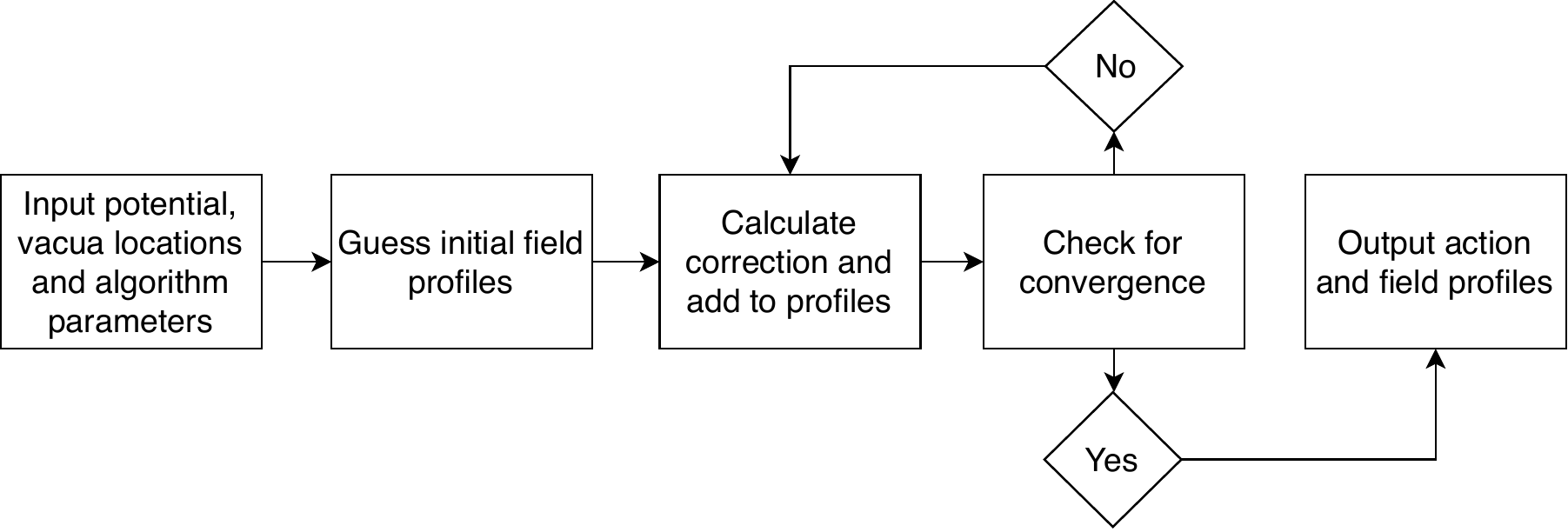}
\caption{Schematic showing the high level structure and execution flow of \bp. A typical execution begins with the user providing a potential, the locations of the true and false vacua, and configuration parameters for the algorithm. These are used to construct an initial profile guess (ansatz). The main execution loop then consists of computing corrections to the ansatz using multiple shooting, adding the correction to the ansatz, and checking for convergence. Once the specified convergence criteria are met, the code outputs the resulting field profiles and Euclidean action. }
\label{fig:code_structure}
\end{figure}

\figref{fig:code_structure} illustrates the perturbative algorithm \bp
uses to find the bounce action when there is more than one field.  In the
following we describe each step, providing a brief summary of how they
are implemented in the code.  \bp uses an object-oriented modular
design with so-called {\it interface} (or {\it abstract}) classes in
many cases, to allow a given component to have more than one
implementation and ensure maximum flexibility and extensibility.  For
a complete reference to \bp, see the API documentation.

\begin{figure}[h]
\centering
\includegraphics[scale=0.7]{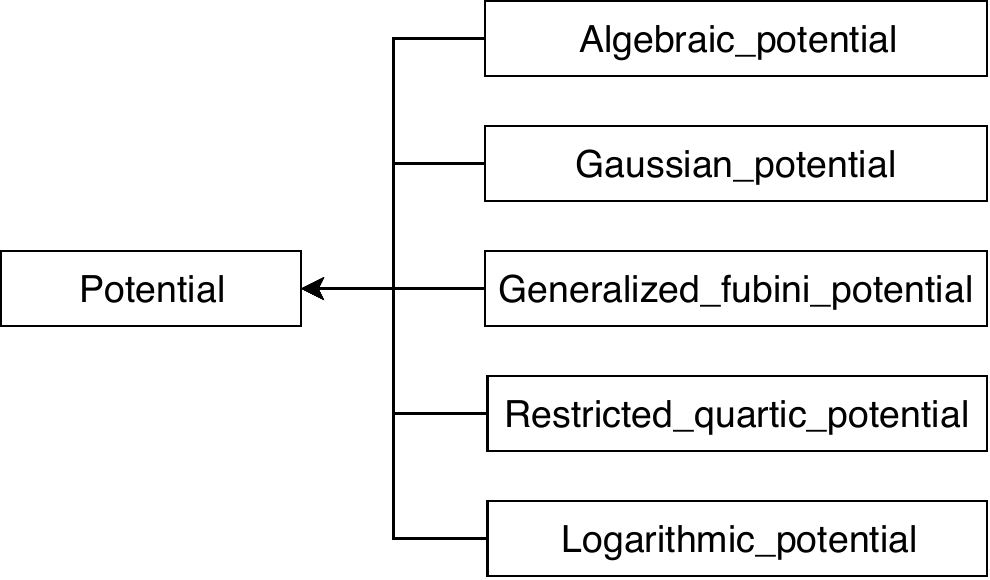}
\caption{Potential class hierarchy.}
\label{fig:classes_potential}
\end{figure}

\begin{figure}[h]
\centering
\includegraphics[scale=0.7]{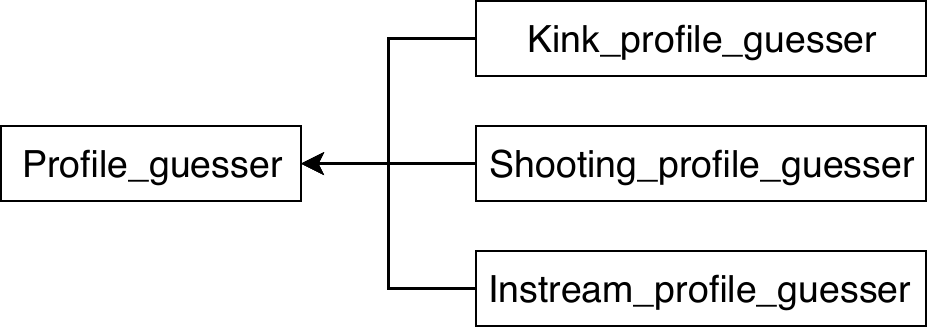}
\caption{Profile guesser class hierarchy.}
\label{fig:classes_profile_guesser}
\end{figure}

\begin{description}
\item[Input potential] An essential input for any bounce solver is the
  potential for which the bounce action must be solved.  \bp contains
  an interface class \code{Potential} for representing real valued
  functions of one or more scalar fields with methods for evaluating
  partial derivatives and making linear changes of coordinates that
  must be given in the inherited class. Specific implementations for
  the potential are then derived from this as shown in
  Fig.\ \ref{fig:classes_potential}. The \code{Algebraic\_potential}
  class is currently used in \bp by default. It uses
  \code{GiNaC}'s~\cite{BauerIntroductionGiNaCFramework2002} symbolic
  manipulation capabilities so that users can input arbitrary
  algebraic expressions as potentials without writing any
  code. This also has the advantage that derivatives and
  changes of coordinates can be implemented at the algebraic level,
  which improves overall performance. This is because in general,
  pre-computing algebraic derivatives will be faster than using
  finite difference methods on each evaluation.
  An additional set of derived classes are provided for testing
  purposes. These are
  \code{Gaussian\_potential}, \code{Generalized\_fubini\_potential},
  \code{Logarithmic\_potential},
  \code{Restricted\_quartic\_potential}, and
  \code{Thin\_wall\_potential}. Full descriptions of these
  potentials may be found in the API documentation. Users who require
  potentials that cannot be expressed via \code{Algebraic_potential}
  are encouraged to implement their own subclasses of \code{Potential}.
  Details of the \code{Potential} interface may be found in the API
  documentation. We encourage users to contact the authors with any
  questions about subclassing \code{Potential}.

\item[Input vacua locations] The purpose of the bounce action solver is to
  find the bounce action for phase transitions between two vacua.
  Therefore a necessary starting point is the location of these
  minima.  The user may specify the locations of both the true and
  false vacua in field coordinates. If the false vacuum location is
  not supplied, the code assumes it is located at the origin. If the
  true vacuum is not supplied, the code will attempt to locate a
  global minimum using the \code{NLopt} optimization
  library~\cite{NLopt}.

\item[Algorithm parameters] A number of options that control the
  execution of the algorithm may be specified. If using the command
  line interface, these are configured using the options specified in
  \appref{sec:User-Options}. If using the code directly, the primary
  configuration points are the \code{Generic_perturbative_profiler} class
  and the associated \code{Profile_convergence_tester} using the setter
  methods described in the API documentation. Available parameters include:
  \begin{itemize}
    \item The domain boundaries, $\rhomin$ and $\rhomax$. If these are not
      specified, the code will attempt to determine appropriate values based
      on the initial ansatz. The domain start is estimated by finding the
      point closest to the origin where the radial derivative of the ansatz
      field profile is equal to $10^{-5}$, while the domain end
      is chosen to be the outermost point at which the ansatz field profile
      is less than or equal to $10^{-5}$.
    \item The integration algorithm. Currently, Runge-Kutta (RK4) and the
      Euler method are supported.
    \item Discretization parameters for integration and interpolation,
      namely the step-size and the fraction of those points that
      are used in spline representations.
   
    \item Tolerances for testing whether the algorithm has converged.
      These can be specified in terms of the Euclidean action, the field values
      at $\rhomin$, or both.
    \item Maximum number of iterations. If the convergence tests do not pass
      before this threshold, an error is issued.
    \item Whether to use the perturbative algorithm or the direct shooting
      method for single field problems.
    \end{itemize}

\item[Guess initial field profiles] In the first step of the \bp
  algorithm the above inputs -- potential, vacua locations, and
  algorithm parameters -- are used to construct an initial set of
  field profiles representing an ansatz solution to the bounce
  equations in \refeq{eq:n_dim_bounce}. To ensure flexibility and
  extensibility, we allow multiple options for constructing the ansatz
  by defining an abstract base class \code{Profile_guesser}. The
  different types of ansatz are implemented as derived subclasses, as
  shown in \figref{fig:classes_profile_guesser}. The two main subclasses are
  \code{Kink_profile_guesser}, which uses the parametric ansatz form given by
  \refeq{eq:kink_ansatz}, and \code{Shooting_profile_guesser} which applies the
  direct shooting method of \secref{sec:shooting} to the reduced one-dimensional
  potential connecting the vacua. \code{Kink_profile_guesser} is the default
  and is slightly faster as it does not involve numerical integration.
  \code{Shooting_profile_guesser} may be of use in unusual cases where the kink
  ansatz does not describe a good initial bubble profile. In either case, the
  resulting set of field profiles is an object from the \code{Field\_profiles}
  class, which uses the \code{eigen3} fast linear algebra library~\cite{eigen3} to
  store a (number of fields) $\times$ (number of grid points) discrete
  representation of the bubble profile. As this array only stores a finite
  number of points, the class also uses the GSL library \cite{GSL} to
  build cubic spline interpolants. This allows off-grid
  evaluation and fast calculation of derivatives using algebraic rather than finite
  difference methods. Finally, \code{Instream\_profile\_guesser} allows the user
  to provide a text file or other input stream containing an ansatz solution.
  This was developed for testing purposes, but may be of use for
  difficult potentials if there are convergence problems with the
  standard ansatz methods. Users may also implement their own subclasses of
  \code{Profile_guesser} for problems where the provided ansatz types are not
  sufficiently close to the true solution \footnote{Typically problematic cases are
  those where the true solution has a high degree of curvature, deviating greatly
  from the straight line bounce paths used by \code{Kink_profile_guesser} and
  \code{Shooting_profile_guesser}.}.

\item[Calculate corrections and add perturbative correction] An
  iteration is then performed to perturbatively find the bounce action
  and field profile. At each step in the iteration a perturbative
  correction is calculated, and used to update the field profile.
  Given a set of field profiles containing an ansatz or partial
  solution, \bp computes a correction function
  $\bm{\epsilon}$ by solving the linearized perturbation
  \refeqs{eq:epsilon_eom}. The \code{Generic\_perturbative\_profiler}
  class accomplishes this via multiple shooting method detailed in
  \secref{sec:multiple_shooting}. Note that as implemented, this
  class is the main entry point for the code, and orchestrates the
  iterative process of successively correcting the profiles until the
  convergence criteria are met. It is \textit{generic} in the sense
  that the user must provide implementations of key dependencies such
  as the convergence tester, ODE integration algorithm, and profile
  guesser. This dependency injection design strategy is intended
  to facilitate flexibility and interchangeability of components.

\item[Convergence check] After each
  perturbative correction to the field profiles, a check is performed
  to determine if the algorithm has converged. This check is
  implemented in the \code{Relative_convergence_tester} class. Objects from this class are stateful and keep track of the Euclidean action $S$ and
  starting field values $\phi_i(0)$ at each iteration. Once the
  relative changes (defined in \refeq{eq:convergence_action}
  and \refeq{eq:convergence_fields}) fall below the corresponding
  thresholds \code{rtol_action} and \code{rtol_fields}, the algorithm
  terminates.

\item[Output] Once the algorithm has converged, the
  \code{Generic_perturbative_profiler} stores the completed
  \code{Field_profiles} and Euclidean action, and makes these available via
  method calls \code{get_bubble_profile} and \code{get_euclidean_action}.

\end{description}

The direct shooting method that can be used for single field problems
has the same inputs and outputs as shown in
\figref{fig:code_structure}, though the specific Algorithm parameters
that are used differ, see \appref{sec:1D_shooting_options} for options
specific to the single-field case.  However the iteration shown in
\figref{fig:code_structure}, which connects the inputs and outputs, is
replaced with the shooting method described in
\secref{sec:shooting}. This is mostly self-contained in the
\code{Shooting} class and sufficient details about the code for this
have already been given in \secref{sec:shooting}, therefore we omit
further details on this here.

%%\section{Examples and test cases}
\section{Comparisons with existing codes and analytic solutions}
\label{sec:tests}
In this section we compare the robustness and performance of our new
code, \bp, comparing to analytic results and against existing
approaches, \cosmo, written in \code{python} and \ab, written in
\code{Mathematica}.  Before presenting the comparisons we will briefly
describe the methods \cosmo and \ab use to find the bounce
solution. To test the accuracy and correctness we will look at the
action calculated by \bp for a special set of potentials where there
are known analytic solutions. We will then look at a more general 1-d
potential where there is no analytic solution and compare \bp to
\cosmo, and investigate the performance of these codes.
Finally we will compare \bp to \ab for multi-field potentials with those of
\cosmo and \ab. All tests were executed on a desktop system running
Ubuntu 16.04, equipped with an Intel i7-4790 3.60 GHz processor and 16
GB of DDR3 RAM clocked at 1.6 GHz.

\subsection{\cosmo and \ab}

For one-field cases \cosmo implements a shooting method in
\code{Python}, which is similar to the direct shooting method
implemented in \bp.  For potentials with more than one field
\cosmo\cite{Wainwright:2011kj} uses a so-called path-deformation
algorithm. The path $\vec\phi(\rho)$ is rewritten in intrinsic
coordinates, parameterized by the distance along the path, $x$. The
equation of motion separates into two pieces
\begin{align}
\ddot x +\tfrac{n}{\rho} \dot x &= \frac{\partial V[\vec \phi(x)]}{\partial x},\\
\frac{d^2 \vec{\phi}}{dx^2} \, \dot x^2 &= \vec\nabla_\perp V(\vec\phi).
\end{align}
The first equation describes the motion along the path whereas the second describes normal forces along the path. For a solution, the normal force,
\begin{equation}
\label{eq:normal_force}
\vec N \equiv \frac{d^2 \vec{\phi}}{dx^2} \, \dot x ^2 - \vec\nabla_\perp V(\vec\phi),
\end{equation}
must vanish such that the second equation is satisfied. In the path-deformation algorithm, the motion along the trajectory, $x(\rho)$, is solved with a shooting method, and the path $\vec\phi(\rho)$ is perturbed by $\Delta\vec\phi(\rho) \propto \vec N(\rho)$. This process is iterated until:
\begin{equation}
\label{eq:ct_stopping_criteria}
\frac{\vec{N}_{max}}{|\nabla V|_{max}} < \kappa ,
\end{equation}
where $N_{max}$ and $|\nabla V|_{max}$ are respectively the largest values of the normal force (equation \ref{eq:normal_force}) and potential gradient along the bounce path, and $\kappa$ is a configurable parameter \footnote{ $\kappa$ corresponds to the \code{fRatioConv} parameter in \cosmo.}.

\ab\cite{Masoumi:2016wot}, on the other hand, uses a multiple shooting method.
The time domain is divided into $n$ subdomains, with boundaries at
$\rho_0,\,\rho_1,\,\ldots\text{ and }\rho_n$. The states at the beginning of
each subdomain, that is $\phi(\rho_i)$ and $\dot\phi(\rho_i)$, are unknown, but
must match the final states of the previous subdomains. They are matched using
Powell's hybrid method. By stitching together solutions in each subdomain, the
method finds a solution for the whole domain.

%% Various alternative methods are proposed in the literature to find the bounce
%% action.  Using that the action has a saddle point at the bounce solution,
%% which is a maximum with respect to dilations, \refcite{Claudson:1983et}
%% extremizes dilations of the action appropriately to find the action at the
%% bounce.  The authors of Refs. \cite{Kusenko:1996jn, Moreno:1998bq} and
%% \cite{John:1998ip} use optimization methods, by defining a minimization
%% function that describes departures from a modified action.
%% \refcite{Kusenko:1995jv} also minimizes a modified action.
%% Refs. \cite{Cline:1999wi, Dasgupta:1996qu} split the equation of motion into
%% two pieces along the lines of \refcite{Wainwright:2011kj}.
%% The authors of Refs. \cite{Cline:1999wi} and \cite{Cline:1998rc} use a gradient
%% ascent/descent method to find the bounce.
%% In \refcite{Konstandin:2006nd} the problem is solved on a lattice.
%% \refcite{Guada:2018jek} connects linear solutions in \refeq{eq:zero}  by approximating the potential by a polygon.
%% \refcite{Espinosa:2018hue} generalizes the single field, thin-wall case by introducing a tunneling potential that connects smoothly the false and true vacua.  The bounce action then is expressed as a simple integral of the tunneling potential.
%% Finally, for the single field case, machine learning techniques are used to find the bounce in \refcite{Jinno:2018jov}.

\subsection{Thin-wall} \label{sec:thin_wall}
\begin{figure}
    \centering
    \begin{subfigure}[b]{0.49\textwidth}
      \includegraphics[width=\textwidth]{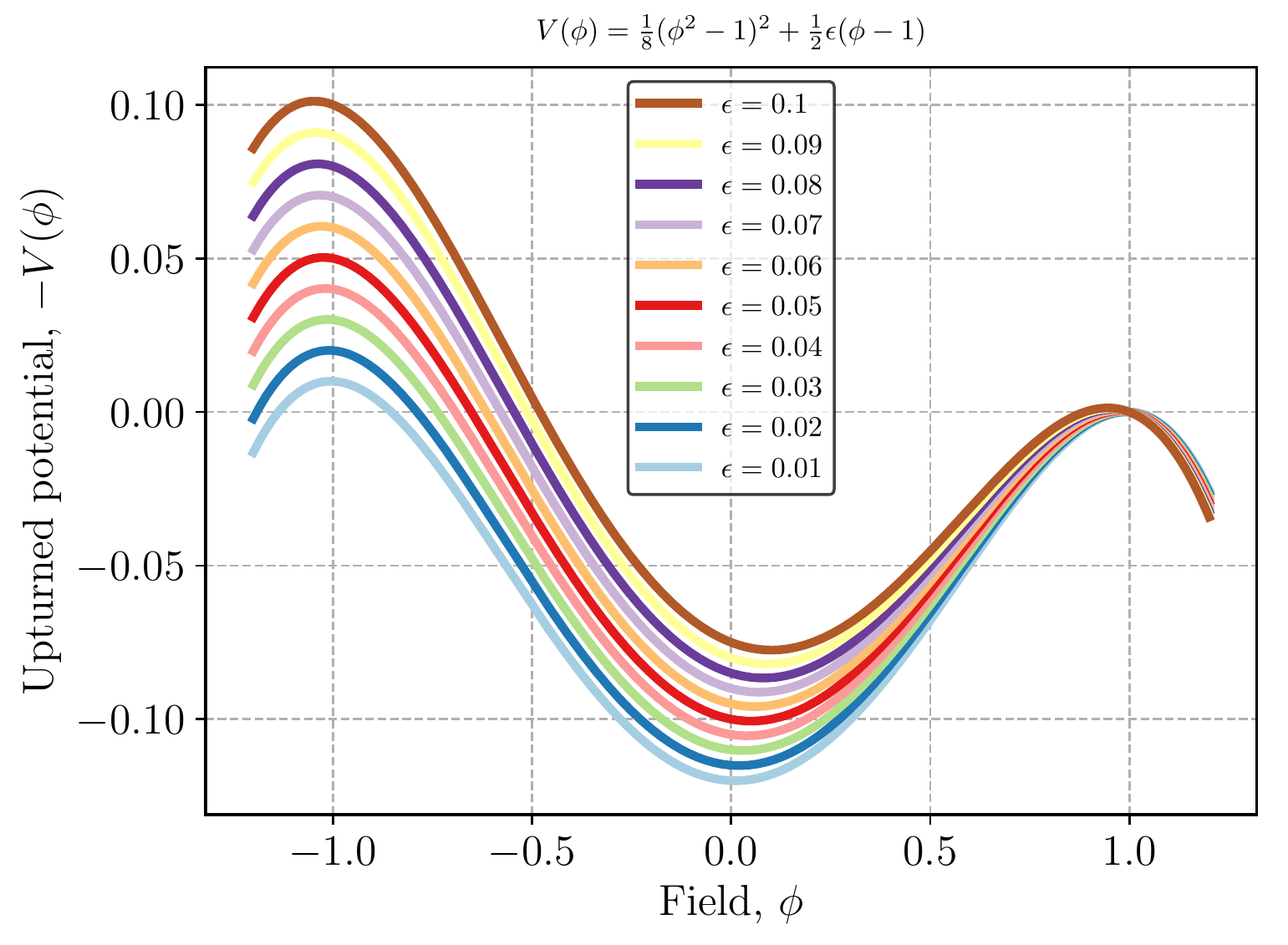}
      \caption{Potential}
      \label{fig:thin_wall_potential}
    \end{subfigure}
    \begin{subfigure}[b]{0.49\textwidth}
      \includegraphics[width=\textwidth]{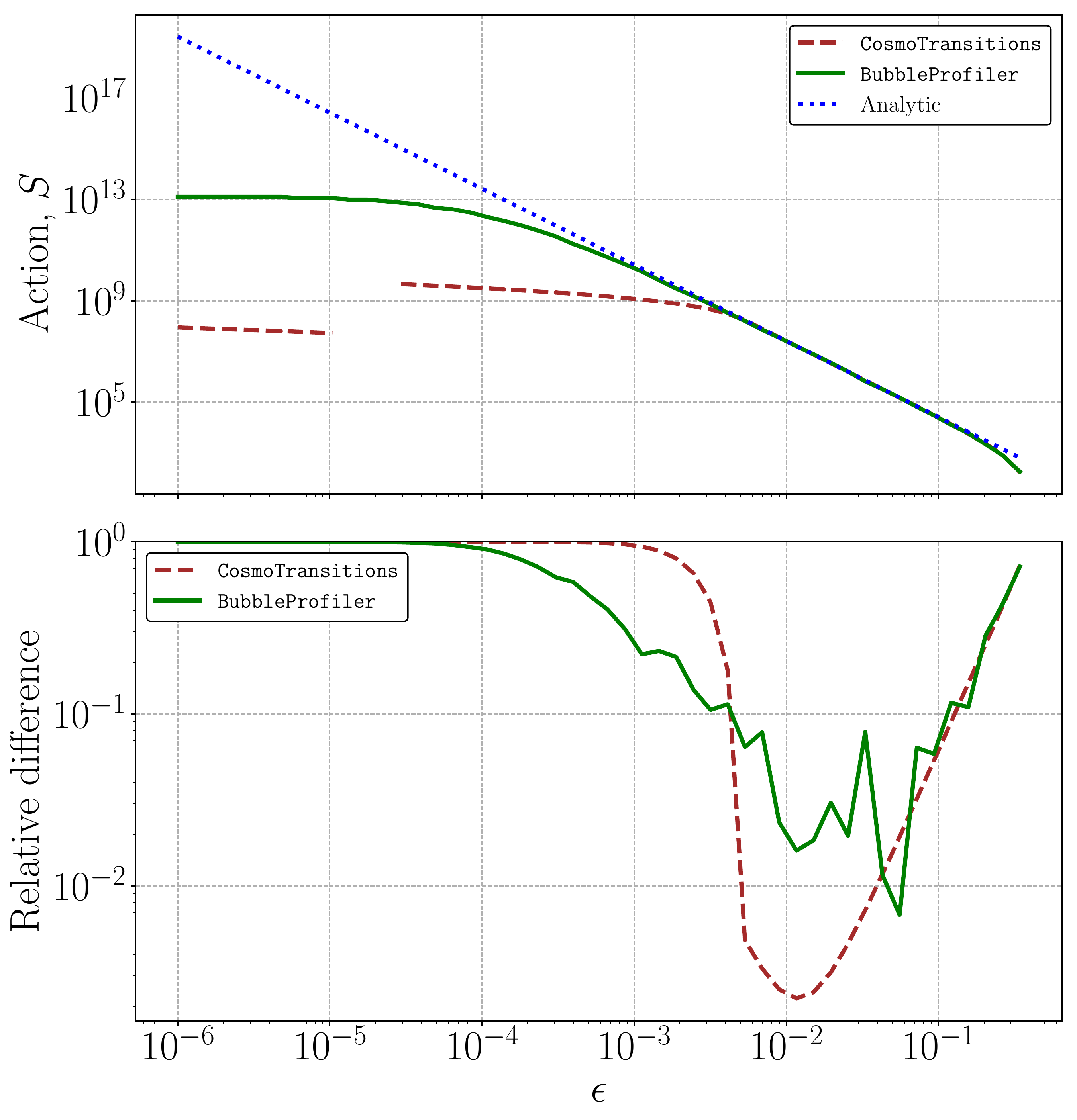}
      \caption{Action}
      \label{fig:thin_wall_action}
    \end{subfigure}
    \caption{\subfigref{fig:thin_wall_potential} Thin-wall potential in
      \refeq{eq:thin_wall}.
      \subfigref{fig:thin_wall_action} Euclidean action in $d=4$ for small values 
      of $\epsilon$. The top panel compares the analytic solution to \bp and
      \cosmo. The bottom panel shows the relative difference between each code and
      the analytic solution. In the region where the bounce solutions from both codes agree well with the thin wall approximation, \bp is generally more accurate than \cosmo. However, in the limit of small $\epsilon$ the action diverges and both codes fail to match the analytic formula.  }
    \label{fig:thin_wall}
\end{figure}

As discussed in \secref{sec:shooting}, when the true and false vacua
of a single-field potential are nearly degenerate, the bounce solution
describes a thin-walled bubble. This poses numerical difficulties for
the direct shooting method, as the system becomes extremely sensitive
to initial conditions. Fortunately, the reparameterization
\refeq{eq:thin_wall_reparam} solves this problem in \bp and \cosmo
also makes use of this trick.  Therefore in this section we
investigate how well these codes reproduce a known thin-wall solution
\cite{Coleman:1977py}.

For $\epsilon / (a^2 \lambda) \ll 1$, the potential,
\begin{equation}\label{eq:thin_wall}
V(\phi) = \frac{\lambda}{8} (\phi^2 - a^2)^2 + \frac{\epsilon}{2 a} (\phi - a),
\end{equation}
has extrema at $\tv \approx -a$, $\barrier \approx \epsilon / (a^3 \lambda)$
and $\fv \approx a$. The thin-wall solution \cite{Coleman:1977py} is,
\begin{equation}
S = \frac{8 \pi^2 a^{12} \lambda^2}{3 \epsilon^3}.
\end{equation}
for $n = 3$\footnote{We added a factor of 16 that was absent in
  \refcite{Coleman:1977py}.} and
\begin{equation}
S = \frac{128 \pi a^9 \lambda^{3/2}}{81 \epsilon^2}
\end{equation}
for $n = 2$, where $n$ appears in \refeq{eq:one_dim_bounce} and can be
related to the number of space-time dimensions, $d$, of the action
which is being calculated, through $n=d-1$.

We checked our code for this potential for $d=4$ in
\code{examples/thin-wall/thin_wall.cpp}, which is built by \code{make
  thin} and executed by \code{bin/thin.x <lambda> <a> <epsilon>}. In
Fig.\ \ref{fig:thin_wall_potential} we show the how the shape of the
potential, for fixed $\lambda = a = 1$, changes as $\epsilon$ is
varied in discrete steps between 0.1 and 0.01, illustrating how the
the thin walled limit is approached as $\epsilon \rightarrow 0$.
Fig.\ \ref{fig:thin_wall_action} then compares the analytic solution
to \bp and \cosmo, varying $\epsilon$ in the range $(10^{-6},
10^{-0.35})$, again fixing $\lambda = a = 1$.  Reading
Fig.\ \ref{fig:thin_wall_action} from right to left one can see that
initially we are away from the thin-walled limit, but as $\epsilon$
is decreased \bp and \cosmo approach the thin-walled, limit reproducing
the analytic solution to within $10^{-1}$ or $10^{-2}$ respectively.
However as $\epsilon$ is reduced further \bp and \cosmo begin to
again diverge from the thin-walled limit, showing that sufficiently
thin-walled cases are still problematic as one would would expect.

\subsection{Fubini potential} \label{sec:fubini}

\begin{figure}
    \centering
    \begin{subfigure}[b]{0.49\textwidth}
      \includegraphics[width=\textwidth]{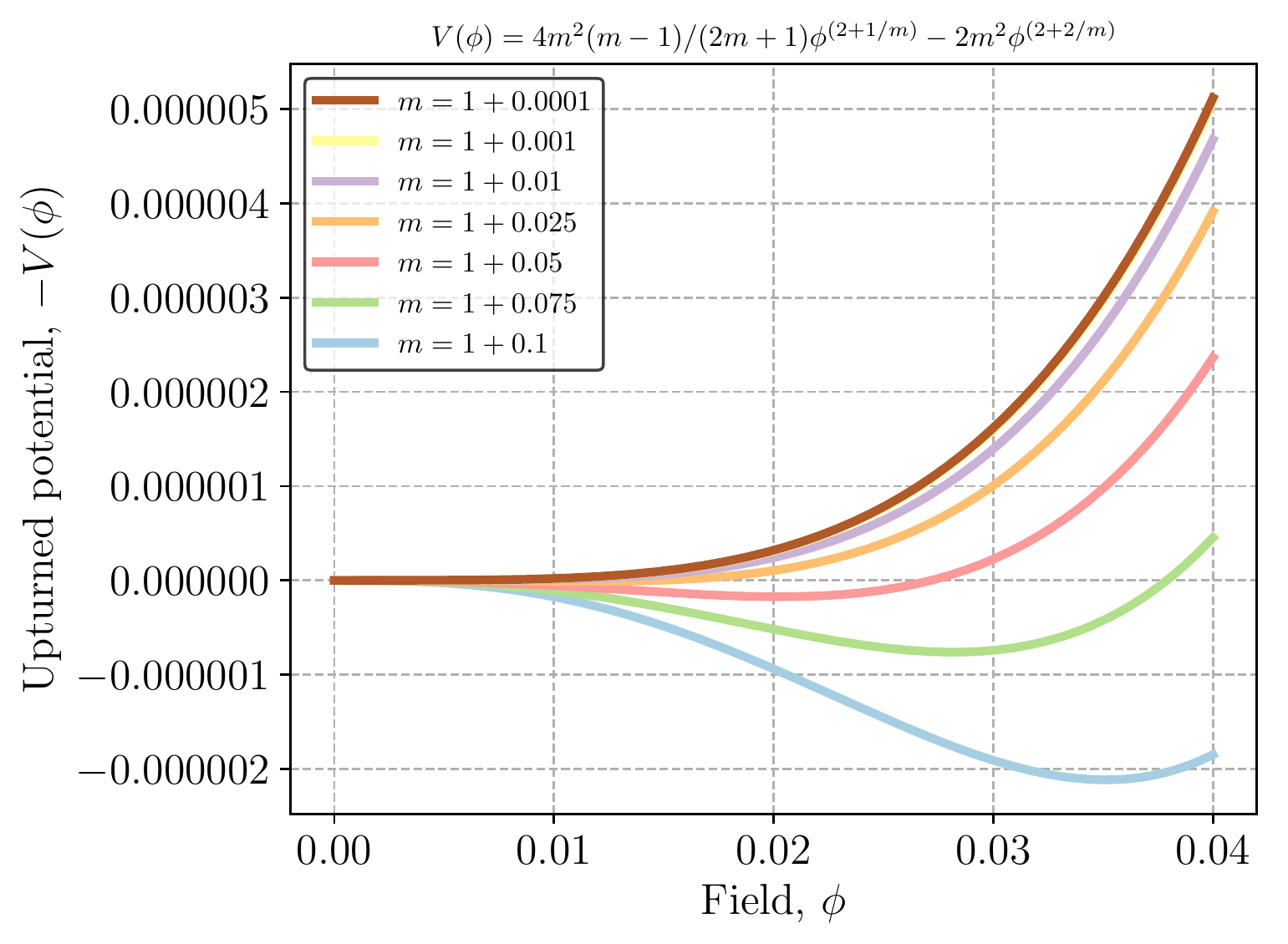}
      \caption{Potential}
      \label{fig:fubini_potential}
    \end{subfigure}
    \begin{subfigure}[b]{0.49\textwidth}
      \includegraphics[width=\textwidth]{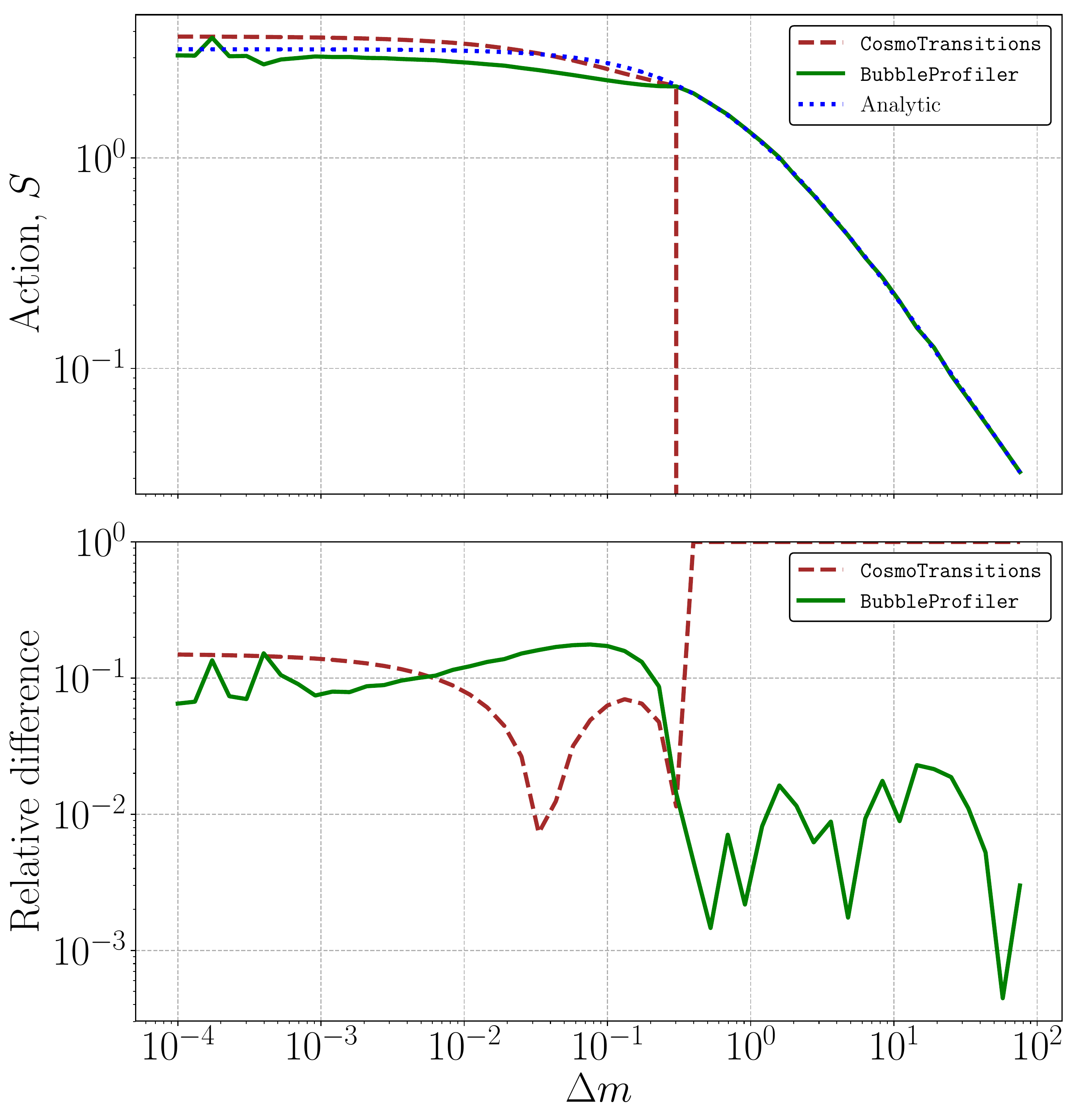}
      \caption{Action}
      \label{fig:fubini_action}
    \end{subfigure}
    \caption{\subfigref{fig:fubini_potential} Fubini potential in
      \refeq{eq:fubini}.
    \subfigref{fig:fubini_action} Euclidean action of the analytic solution, \bp, and \cosmo with absolute values (top) and relative difference (bottom).}
    \label{fig:fubini}
\end{figure}

There is also a known analytic result for the generalized Fubini potential,
\begin{equation}\label{eq:fubini}
V(\phi) = \frac{4 u m^2 (m-1)}{2 m+1}\phi^{(2m+1)/m} - 2 u v m^2 \phi^{(2m+2)/m},
\end{equation}
which results in the bounce action \cite{Aravind:2014pva}
\begin{equation}
S = \frac{m \pi^2}{(4 m^2-1)}\frac{1}{uv^{2m-1}},
\end{equation}
for $n=3$. We checked our code for this potential in
\code{examples/general-fubini/general_fubini.cpp}, which is built by
\code{make fubini} and \code{bin/fubini.x <u> <v> <m>}.  We compared
the analytic solution to \bp and \cosmo using $u = v = 1$, and varying
$m = 1 + \Delta m$ in the range $\Delta m = (10^{-4},10^{2})$. The
results are summarized in \figref{fig:fubini}, where we again
illustrate how the shape of the potential changes on the left hand
side.  On the right hand side we present results for the action for
\bp, \cosmo and the analytic solution in the top panel and in the
bottom panel show the relative difference between the analytic
solution and the action calculated by the two codes. For $\Delta m <
10^{-1}$, the accuracy of the two codes is comparable, though both codes struggle with precise agreement
with the analytic solution. This is because as $\Delta m \to 0$ the barrier vanishes
resulting in an infinitely thick-wall and consequently numerical problems.
For $\Delta m >
1$, \cosmo fails to produce a solution, while \bp matches the analytic
solution to within a tolerance of $10^{-1}$.

\subsection{Logarithmic potential} \label{sec:logarithmic}

\begin{figure}
    \centering
    \begin{subfigure}[b]{0.49\textwidth}
      \includegraphics[width=\textwidth]{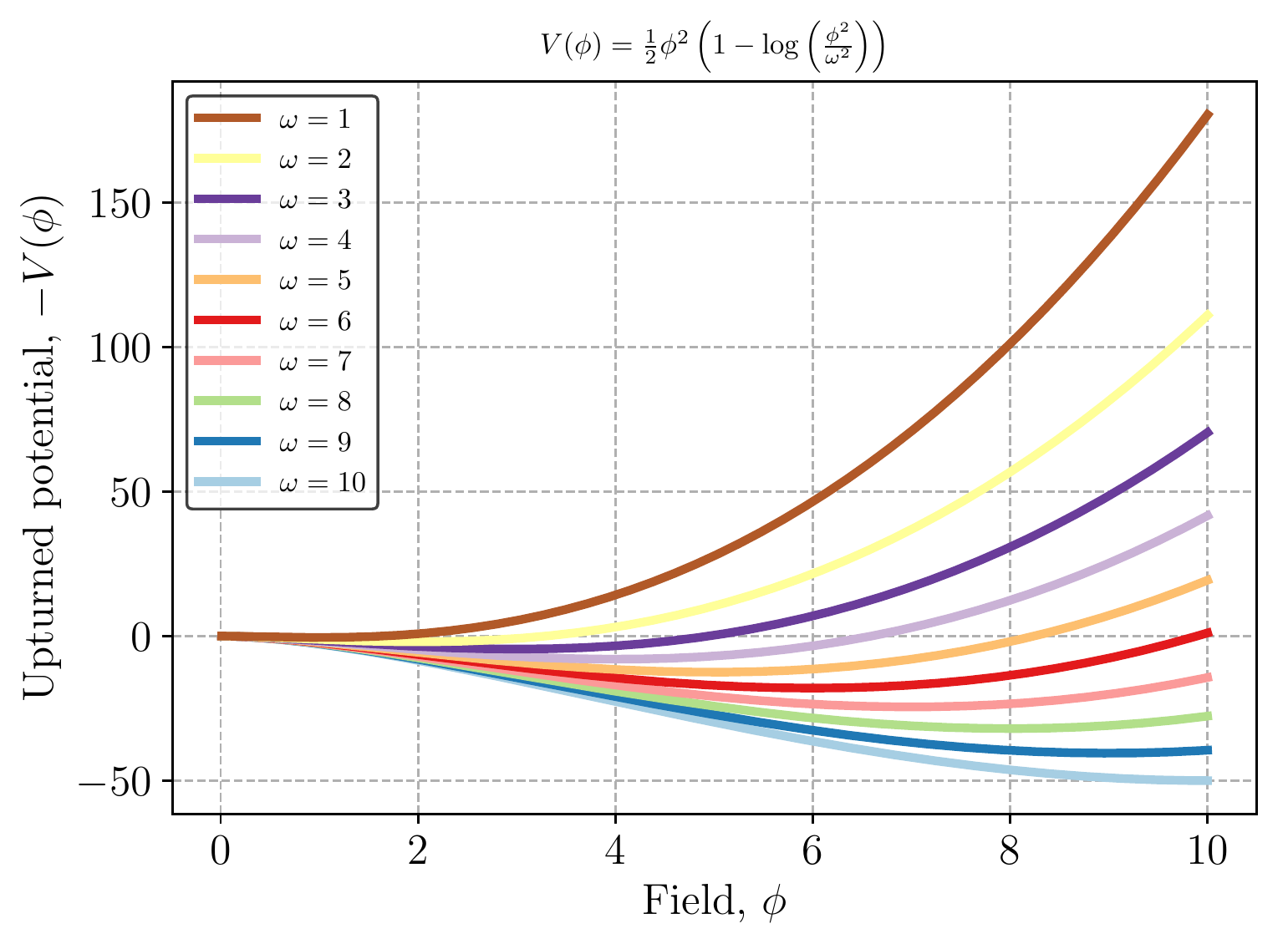}
      \caption{Potential}
      \label{fig:logarithmic_potential}
    \end{subfigure}
    \begin{subfigure}[b]{0.49\textwidth}
      \includegraphics[width=\textwidth]{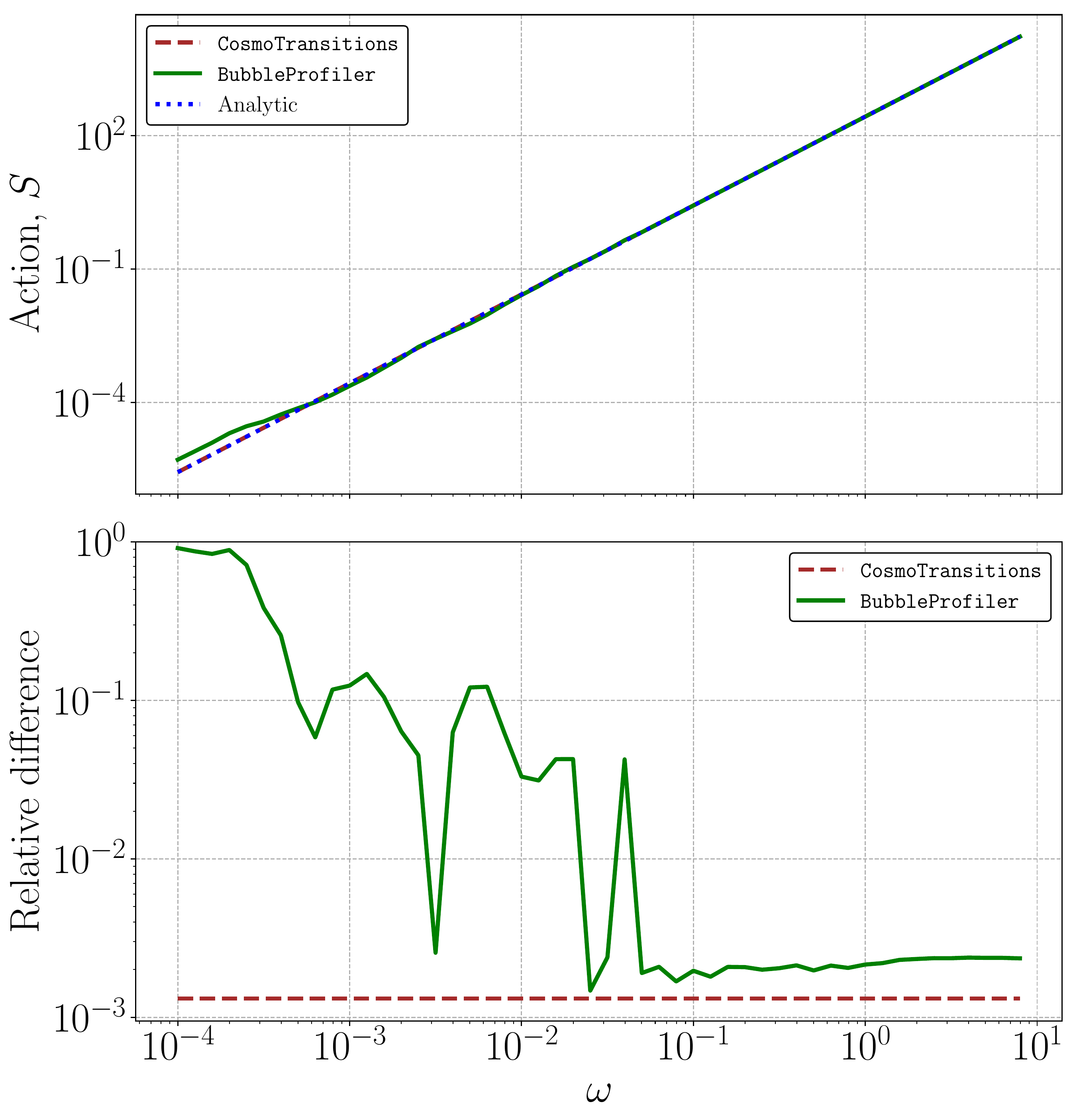}
      \caption{Action}
      \label{fig:logarithmic_action}
    \end{subfigure}
    \caption{\subfigref{fig:logarithmic_potential} Logarithmic potential in
      \refeq{eq:logarithmic}.
    \subfigref{fig:logarithmic_action} Euclidean action of the analytic solution, \bp, and \cosmo with absolute values (top) and relative difference (bottom). }
    \label{fig:logarithmic}
\end{figure}

Another potential with known analytic solutions is the logarithmic
potential,
\begin{equation}\label{eq:logarithmic}
V(\phi) = \frac12 m^2 \phi^2 \left[1 - \ln \left(\frac{\phi^2}{w^2}\right) \right],
\end{equation}
and this results in the bounce action \cite{Aravind:2014pva}
\begin{equation}
S = \frac{\pi^2 e^4}{2}\frac{w^2}{m^2},
\end{equation}
for $n=3$. We checked our code for this potential in
\code{examples/logarithmic/logarithmic.cpp}, which is built by
\code{make logarithmic} and executed by \code{bin/logarithmic.x <m> <w>}.  We
compared the analytic solution to \bp and \cosmo using $m = 1$ and varying
$\omega$ in the range $(10^{-4}, 10^1)$. The results are summarized in
\figref{fig:logarithmic}. We found that for this potential, \cosmo was more
accurate with the difference being most pronounced for low values of $\omega$.

\subsection{Renormalizable single-field potentials}\label{sec:reparam}

We now consider potentials of the form
\begin{equation}\label{eq:1d_E_alpha}
V(\phi) = \frac{-4 \alpha +3}{2} \phi^2-  \phi ^3 + \alpha \phi ^4.
\end{equation}
Solutions to arbitrary order-four polynomials are related by rescalings to
solutions to this restricted order-four polynomial, as we will now describe.

\begin{figure}
    \centering
   %% \begin{subfigure}[b]{0.49\textwidth}
      \includegraphics[width=0.75\textwidth]{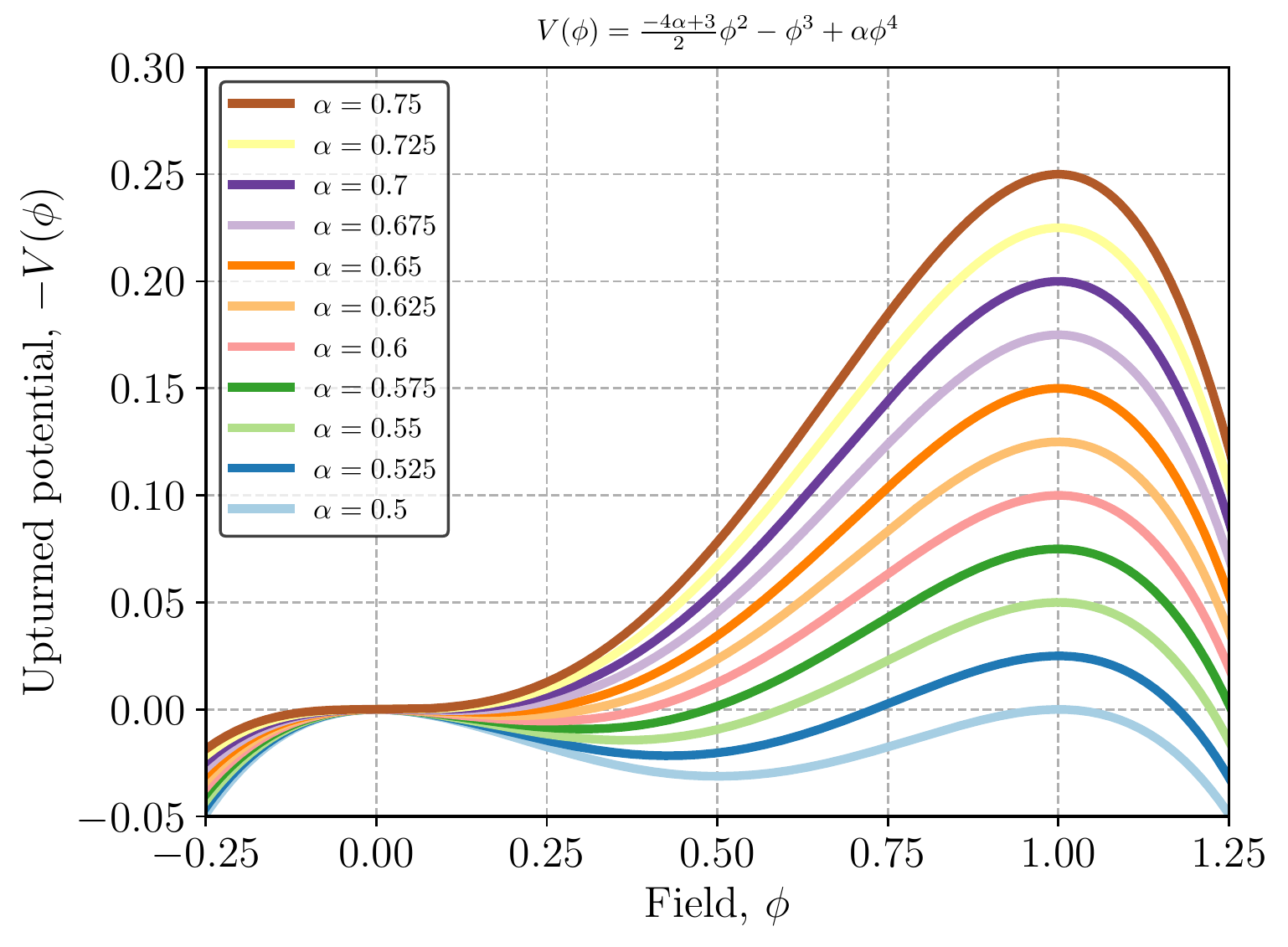}
    \caption{ Quartic potential in
      \refeq{eq:1d_E_alpha} for $\alpha = (0.5, 0.75)$.  The parameter
      $\alpha$ smoothly interpolates between thin and thick-walled
      potentials.  For $\alpha \simeq 0.5$, the true vacuum, $\tv =
      1$, and the false vacuum, $\fv = 0$, are degenerate and the
      bounce is thin-walled, whereas for $\alpha \simeq 0.75$, the
      barrier and the false minimum nearly coincide and the solution
      is thick-walled.}
    \label{fig:quartic_potential}
\end{figure}

Any potential with two minima separated by a local maxima can be approximated by a quartic potential, which naively has five free parameters. However, four of these parameters are redundant as bounce solutions to classes of potentials related by trivial transformations are themselves related by trivial transformations. Specifically, the bounce action is invariant under the transformations $\phi \to \phi + \Delta\phi$ and $V \to V + \Delta V$, eliminating two parameters. The transformation $V \to V / a$ changes the action by $S \to \sqrt{a} S$ for $d=3$ and by $S \to a S$ for $d=4$, and eliminates a further parameter. Finally, we may work in units of our choice: the changes of unit, $\phi \to \phi / b$ etc, result in $S \to S / b$ for $d=3$, and leave the action unchanged in $d=4$ as it is dimensionless.

This leaves a single parameter relevant for solving the bounce equation and thus we may make the problem a single parameter problem. Beginning with a general renormalizable potential,
\begin{equation}
\label{eq:general_quartic_potential}
V(\phi) = \Lambda^4 + t \phi + m^2 \phi^2 + \kappa \phi ^3 + \lambda \phi^4,
\end{equation}
with true vacuum $\tv$ and false vacuum $\fv$, we shift the potential such that the false vacuum is at the origin and pick units such that the true vacuum is at $\phi= 1$. We rescale the potential such that the coefficient of the cubic term is minus one and remove the constant piece. We are left with a potential described by a single parameter $\alpha$,
\begin{equation}\label{eq:one_dim_simplified}
V(\chi) = \frac{-4 \alpha +3}{2} \chi^2 - \chi ^3 + \alpha \chi^4,
\end{equation}
where
\begin{equation}
\alpha = \lambda \left|\frac{\tv - \fv}{\kappa + 4 \lambda \fv}\right|.
\end{equation}
By the fact that the true vacuum is at $\chi = 1$ and thus $V(\chi=1) < V(\chi=0)$, and the fact that the origin is a minima, $V^{\prime\prime}(\chi=0) > 0$, we require $1/2 < \alpha < 3 /4$ for consistency.  In \figref{fig:quartic_potential} we plot potentials with $\alpha$ varying over this range in steps of $0.025$.

We denote the action for this potential by $S(\alpha)$. The action for the original, general quartic potential depends on a potential-specific factor that simply scales $S(\alpha)$. For $d=3$,
\begin{equation}
S = |\tv - \fv| \sqrt{\frac{\lambda}{\alpha}} \, S(\alpha).
\end{equation}
and for $d=4$,
\begin{equation}
S = \frac{\lambda}{\alpha} \, S(\alpha).
\end{equation}
Thus it suffices to consider $S(\alpha)$ and potentials parameterized by
$\alpha$. We find from a thin-wall approximation assuming
$\alpha \simeq 1/2$ that for $d=3$,
\begin{equation}\label{eq:action_alpha_thin_3}
S(\alpha) = \frac{2 \pi }{81} \frac{1}{(\alpha - 1/2)^2},
\end{equation}
and for $d=4$,
\begin{equation}
S(\alpha) = \frac{\pi^2}{96} \frac{1}{(\alpha - 1/2)^3}.
\end{equation}

We checked our code for this potential in
\code{examples/quartic/action.cpp}, which is built by \code{make
  quartic} and executed by \code{bin/quartic.x <E> <alpha>
  <dim>}.  The result may be tabulated for $E=1$ by the program
\code{examples/quartic/tabulate.cpp}, which is built by \code{make
  quartic_tabulate} and executed by\newline
\code{bin/quartic_tabulate.x <dim> <min-alpha> <max-alpha> <step>}. The perturbative algorithm is intended for multi-field cases, but we also include it in this analysis for purposes of comparison.

As discussed in \secref{sec:thin_wall} there is a known analytic solution in the thin-walled limit. With the parameterization \refeq{eq:one_dim_simplified}, $\alpha$ controls the degree of degeneracy. In particular, as $\alpha \rightarrow 0.5$, the vacua become degenerate and the bounce solution approaches a step function \cite{Akula:2016gpl}.

\begin{figure}[h!]
\centering
  \begin{subfigure}[b]{0.49\textwidth}
    \includegraphics[width=\textwidth]{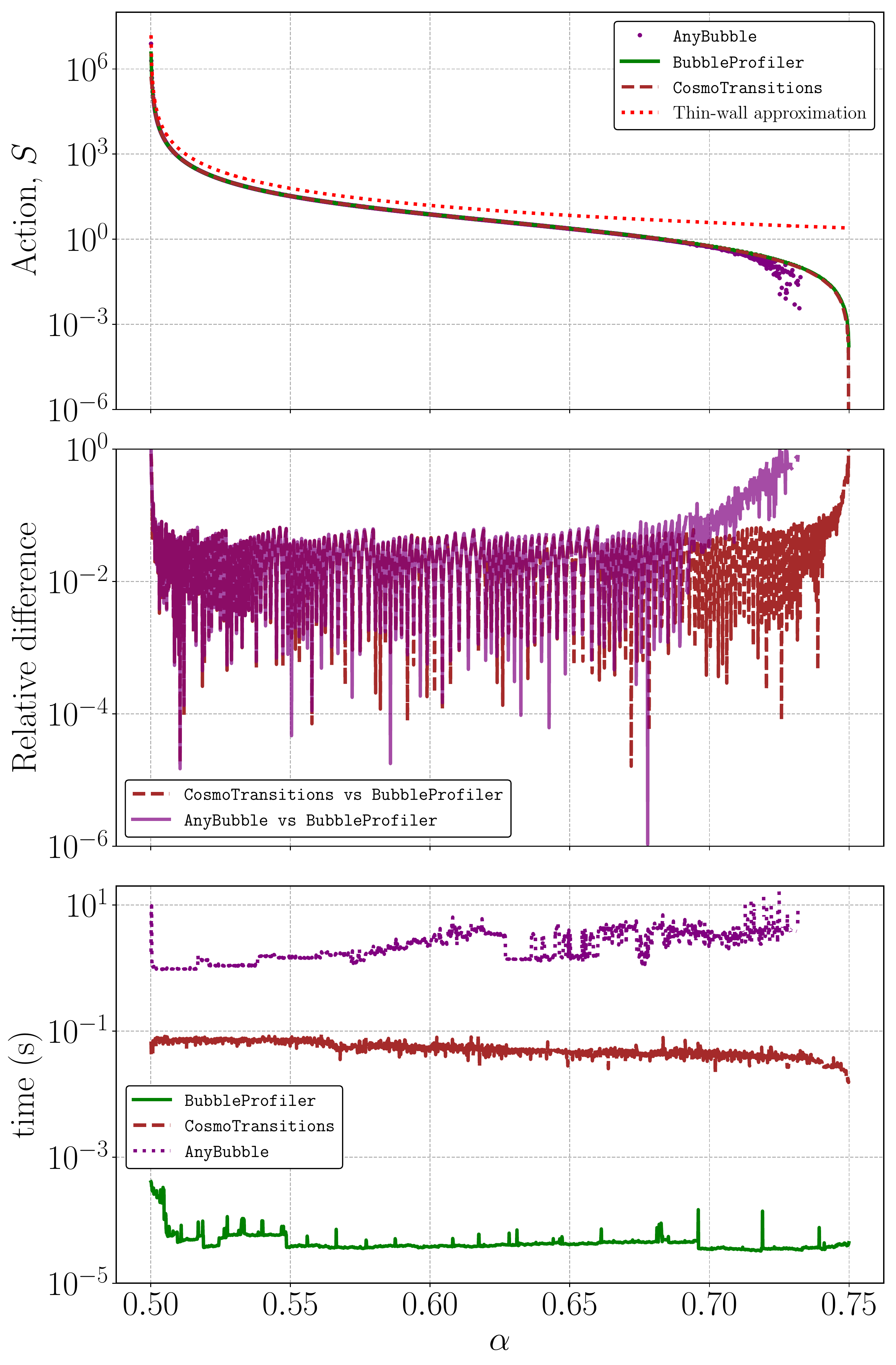}
    \caption{Shooting method}
    \label{fig:alpha_tests_shooting}
  \end{subfigure}
  \begin{subfigure}[b]{0.49\textwidth}
    \includegraphics[width=\textwidth]{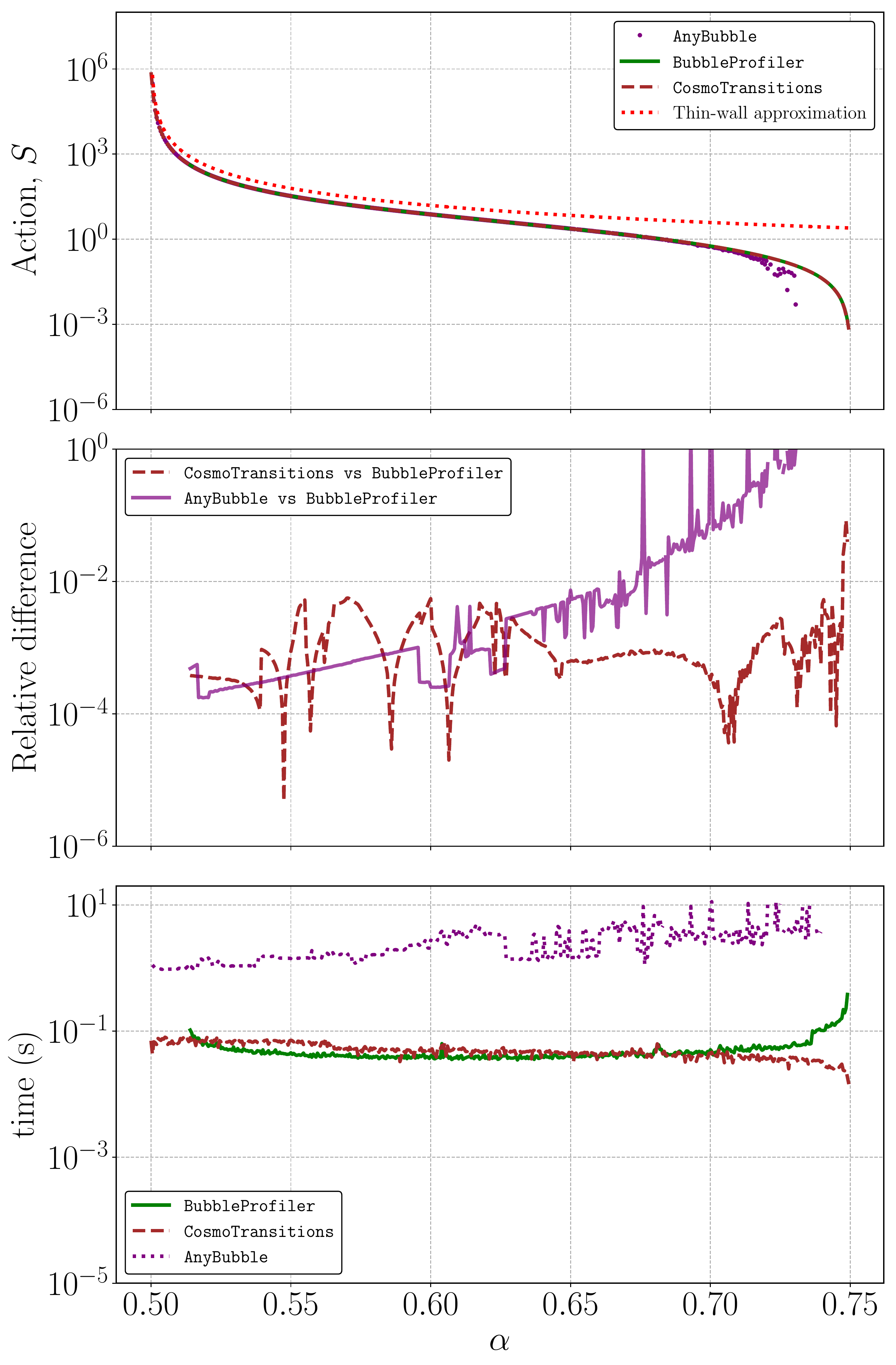}
    \caption{Perturbative algorithm}
    \label{fig:alpha_tests_perturbative}
  \end{subfigure}
\caption{Comparison of \subfigref{fig:alpha_tests_shooting} direct
  shooting with \bp and direct shooting with \cosmo and \ab, and
  \subfigref{fig:alpha_tests_perturbative} the perturbative algorithm
  implemented by \bp with \code{initial-step-size} set to 0.1 and direct
  shooting with \cosmo and \ab. The potential is given by
  \refeq{eq:general_quartic_potential}, with $E = 1$ and $\alpha \in
  (0.5,0.75)$. From top to bottom, the three panels in each chart show
  the Euclidean action, $S_E$; the relative difference in $S_E$
  between \bp and \cosmo and between \bp and \ab; and the execution time.}
\label{fig:alpha_tests}
\end{figure}

\figref{fig:alpha_tests} shows that when using the direct shooting
method, \bp outperforms \cosmo, and --- excepting the asymptotic cases
$\alpha \rightarrow 0.5$ and $\alpha \rightarrow 0.75$ --- computes
the same Euclidean action to within a factor of $10^{-1}$.  Since the
direct shooting part of \bp is a C++ implementation of the algorithm in \cosmo (a \code{Python} code), this is
unsurprising. For single field problems, the speed of the perturbative
algorithm depends strongly on the \code{initial-step-size} setting.
With \code{initial-step-size} set to 0.1, which we used here and in
testing delivered a stable action calculation without degradation of
the precision, we find timings that are quite similar to that of
\cosmo for all $\alpha$ other than those close to the asymptotic
limits.  In fact in this range the agreement between the perturbative
method of \bp and \cosmo is actually slightly higher, with the two
codes agreeing within $10^{-2}$.  However the perturbative method
fails consistently for thin walled cases with $\alpha < 0.51$.  This
thin-walled problem will be discussed further when we consider
multi-field potentials.  For thick walls there is also some
performance degradation and agreement between the codes is a little
worse.
%% , which we discuss further in the next
%% -section.

\subsection{Multi-field potentials}
\subsubsection{Thin walled bubbles in multi-field potentials}
In applying the perturbative algorithm to single field potentials we found that
cases with near-degenerate vacua resulting in thin walled bubbles became
problematic for $\alpha < 0.51$. To investigate whether this problem generalized
to multi-field potentials we implemented the \code{Gaussian_Potential} class.
This class implements a potential of the form:
\begin{equation}
V(\phi) = -(N(\phi, 0) + \gamma N(\phi, \mu)),
\end{equation}
where
\begin{equation}
  N(\phi,\mu) = \frac{1}{(2\pi)^{n/2}} \text{exp}\bigg(-\frac{1}{2}
  |\phi - \mu|^2 \bigg)
\end{equation}
is a unit n-dimensional Gaussian, $\gamma$ controls the relative depth of the
minima, and $\mu = 1/\sqrt{n}(\lambda, ..., \lambda)$ so that $\lambda$ is the
geometric distance between minima. As $\gamma \rightarrow 1$, the vacua approach
degeneracy and the solution becomes thin walled.  We tested potentials of up to
five fields and found that the problem generalized in a straightforward way -
if the value of $\alpha$ corresponding to the single-field potential used to
fit the ansatz approached $0.51$, the profiler failed to converge.

In light of this result, we added a check which compares the ansatz $\alpha$ to
a threshold \code{Kink_profile_guesser::alpha_threshold = 0.514}. An error is
issued if the value is less than the threshold. Note that this check is only
applied when using the perturbative algorithm. For single field potentials, the
shooting method can be used to solve thin walled bounces.

\subsubsection{Multi-field potentials --- comparison with other codes}
\label{multi_field_benchmarks}

For $n > 1$ fields, the direct shooting method of \secref{sec:shooting} is no
longer applicable. We devised a set of non-physical, polynomial potentials of
between one and eight fields for the purposes of comparison. To compare the
performance of all three algorithms, and the extent to which they agree on the
Euclidean action of the bounce solution, each code was run on these test
potentials; the results are summarized in \tabref{multi_field_testing}.  The
potentials used are listed in \appref{app:test_potentials}. These tests are
also included as a script in the \code{bubbler} tool (see \appref{sec:bubbler})
called \code{n_fields_from_interface.py}

The \bp test results were obtained by executing commands of the form: 
\begin{lstlisting}
run_cmd_line_potential.x --force-output --write-profiles --potential <potential> --field <fields> --initial-step-size=0.1 --domain-start -1.0 --domain-end -1.0 --local-minimum 0.0 --local-minimum 0.0 --global-minimum 1.0 --global-minimum 1.0 --rtol-action 0.001 --rtol-fields 0.001 --integration-method runge-kutta-4
\end{lstlisting}
substituting appropriate values of \code{<potential>} and \code{<fields>} for each test.

\begin{table}[h]
\centering
\begin{tabular}{c c c c c c c}
\toprule
& \multicolumn{3}{c}{Action} & \multicolumn{3}{c}{Time (s)} \\
\cmidrule(r){2-4} \cmidrule(r){5-7}
\# fields & BP & CT & AB & BP & CT & AB \\
\midrule
1 & 54.1 & 52.6 & 52.4 & 0.051 & 0.066 & 1.285 \\
2 & 20.8 & 21.1 & 20.8 & 0.479 & 0.352 & 7.473 \\
3 & 22.0 & 22.0 & 22.0 & 0.964 & 0.215 & 25.209 \\
4 & 55.9 & 56.4 & 55.9 & 1.378 & 0.255 & 54.258 \\
5 & 16.3 & 16.3 & 16.3 & 2.958 & 0.367 & 305.531 \\
6 & 24.5 & 24.5 & 24.4 & 4.853 &  0.337 & 830.449 \\
7 & 36.7 & 36.6 & 36.7 & 6.754 & 0.375 & 1430.892 \\
8 & 46.0 & 46.0 & 46.0 & 10.014 & 0.409 & 1805.713 \\
\bottomrule
\end{tabular}
\caption{Test results for potentials of $1$--$8$ fields for \bp (BP),
  \cosmo (CT), and \ab (AB). \bp was configured with
  \code{initial-step-size} set to $0.1$ and a stopping criteria of
  \code{rtol_action = rtol_fields = 0.001} (see
  \secref{sec:perturbative_corrections}). The comparable parameter for \cosmo
  is \code{fRatioConv} (see \refeq{eq:ct_stopping_criteria}), which we set to
  the default value of \code{0.02}. \ab does not have a configurable stopping
  criterion.}
\label{multi_field_testing}
\end{table}

We find that \cosmo is faster than \bp for all cases
other than the single field case, where \bp is slightly faster, though
the differences in speed are much less significant when there are only
a few fields. \ab is significantly slower than the other two codes,
although this may be a reflection of the current \code{Mathematica}
implementation, rather than the underlying algorithm. The degree to
which the codes agree on the Euclidean action varies, but is within
$4\%$ in all cases. If we exclude the single field case, where \bp
uses direct shooting rather than the perturbative algorithm, the codes
agree to within $2\%$.

\section{Scalar Singlet Model}
\label{sec:ssm}
For a realistic application of \bp we now consider a standard
model extension where the Higgs sector has an additional (real) scalar
singlet field.  This scalar singlet model (SSM) is arguably the
simplest possible extension of the standard model of particle
physics. Nonetheless, despite the minimality of the SSM, it has
generated extensive interest\footnote{For the current status of the
  model see recent global fits in
  Refs.\ \cite{Athron:2017kgt,Athron:2018ipf}.} as unlike the SM, it
can both explain the relic density of dark matter
\cite{McDonaldGaugesingletscalars1994,
  BurgessMinimalModelnonbaryonic2001} and, relevant for our work here,
 supports a first order EWPT with the Higgs mass at
$125\,\text{GeV}$
\cite{KurupDynamicselectroweakphase2017,ClineElectroweakbaryogenesisdark2013,Vaskonen:2016yiu}.
As such this provides a relevant and interesting model for us to
illustrate how one can use \bp when investigating the properties of
the EWPT in a realistic model. %% In this section, we present the
%% relevant details and simplifying assumptions necessary to carry out
%% this testing. For a more detailed exposition of the model, see
%% \refcite{EspinosaStrongElectroweakPhase2012}.

The most general renormalizable potential coupling the Higgs scalar
$h$ to the new scalar singlet $s$ depends on eight operators.  We,
however, impose a $\mathbb{Z}_2$ symmetry $s\rightarrow -s$, which
permits only five operators in the potential,
\begin{equation}
V_\text{Tree} = -\frac{1}{2}\mu_h^2 h^2 + \frac{1}{4}\lambda_h h^4 + \frac{1}{2}\mu_s^2 s^2 + \frac{1}{4} \lambda_s s^4 + \frac{1}{4} \lambda_m s^2 h^2.
\end{equation}
Following \refcite{EspinosaStrongElectroweakPhase2012} we add only
the leading order terms obtained from a high temperature expansion of
the one-loop thermal corrections to the potential, which is
sufficient for the scenarios we consider.  This gives,
\begin{equation}
V_T = V_\text{Tree} + \frac{1}{2}(c_h h^2 + c_s s^2) T^2.
\end{equation}
where $c_h$ and $c_s$ are defined by
\begin{align}
c_h &= \frac{1}{48}(9g^2 + 3g^{\prime 2} + 2(6 h_t + 12 \lambda_h + \lambda_m)), \label{eq:c_h} \\
c_s &= \frac{1}{12}(2 \lambda_m + 3 \lambda_s), \label{eq:c_s}
\end{align}
and $g$, $g^{\prime}$ are the weak charge and weak hypercharge
couplings respectively, while $h_t$ is the top quark Yukawa coupling.

At high temperatures the terms from the finite temperature potential
ensure the quadratic terms have positive coefficients and electroweak
symmetry is restored with a global minimum at the origin.  We will
focus on scenarios where as the temperature cools down a deeper
minimum develops in the $h=0$ directions with a non-zero value for
singlet field, $\langle s \rangle = w$, which spontaneously breaks the
$\mathbb{Z}_2$ symmetry.  As the temperature cools further the
electroweak minima with $\langle h \rangle =v$ and $\langle s \rangle
=0$ develops and there is a first order phase transition from the
$\mathbb{Z}_2$ breaking minimum to the electroweak minimum where the
$\mathbb{Z}_2$ symmetry is restored.  Our aim is to calculate the
bounce action for this transition, and we will use this to determine if
bubble nucleation takes place and if so, what the nucleation
temperature is.

First we require that at zero temperature there is an electroweak
symmetry breaking (EWSB) minimum with a vacuum expectation value of
$v_\text{EW} = 246.22$ GeV.  This allows us to use the EWSB condition,
to fix
\begin{equation}
\label{Eq:mu_h_Zero_T}
  \mu_h^2 = \lambda  v_\text{EW},
\end{equation}
where this matches the familiar SM relation, since the
singlet VEV is zero in the EWSB minimum. Secondly we use
the measured value of the Higgs mass, $M_H =125.1$ GeV
\cite{Khachatryan:2016vau} to fix the
quartic Higgs coupling,
\begin{equation}
\label{eq:lambda_h}
\lambda_h = \frac{M_H^2}{2 v^2_{EW}}.
\end{equation}
Next we require that at the critical temperature $T=T_C$ the
electroweak vacuum is degenerate with the $\mathbb{Z}_2$ breaking
minimum.  This will allow us to replace one of the remaining
parameters with the critical temperature. To do this we again follow
\refcite{EspinosaStrongElectroweakPhase2012}, and introduce
temperature dependent quadratic couplings, so that the temperature
dependence of the potential is absorbed into these couplings,
\begin{equation}
\tilde{\mu}_h^2(T) = \mu_h^2 - c_h T^2,\ \tilde{\mu}_s^2(T) = \mu_s^2 + c_s T^2,
\end{equation}
and
\begin{equation}
V_T = -\frac{1}{2} \tilde{\mu}_h^2(T) h^2 + \frac{1}{2} \tilde{\mu}_s^2(T) s^2 + \frac{1}{4} \lambda_h h^4 + \frac{1}{4} \lambda_s s^4 + \frac{1}{4} \lambda_m s^2 h^2.
\end{equation}
We can now investigate the minima at finite temperature. Taking the first derivative with respect to the $h$ field,
%% \begin{equation}
%% \frac{\partial V_T}{\partial h} = - \tilde{\mu}^2_h(T) h + \lambda_h h^3 + \frac{1}{2} \lambda_m s^2
%% \end{equation}
%
and evaluating at the symmetry breaking vacuum, where it vanishes, leads to,
\begin{equation}
\label{eq:higgs_ewsb}
v^2(T) = \frac{\tilde{\mu}_h^2(T)}{\lambda_h} = v^2_\text{EW} - \frac{c_h}{\lambda_h} T^2.
\end{equation}
%% Note \refeq{eq:higgs_ewsb} implies
%% $\tilde{\mu}_h^2(T=0) = \mu_h^2 = \lambda_h v^2_1{EW}$ in agreement
%% with \refeq{Eq:mu_h_Zero_T} as one would expect.

We obtain a  singlet mass at finite temperature through,
\begin{equation}
m_s^2(T) = \frac{\partial^2 V_T}{\partial s^2}\bigg|_{\substack{h = v\\s = 0}} = \tilde{\mu}_s^2(T) + \frac{1}{2} \lambda_m v^2(T)
\label{eq:singlet_mass}
\end{equation}
\refcite{EspinosaStrongElectroweakPhase2012} shows that imposing degeneracy of the symmetric and symmetry breaking vacua at $T = T_C$ results in the constraint:
\begin{equation}
m_s^2(T_C) = \frac{v^2(T_C)}{2}(\lambda_m - 2 \sqrt{\lambda_h \lambda_s}).
\end{equation}
Inserting this into \refeq{eq:singlet_mass} and rearranging gives,
\begin{equation}
\tilde{\mu}_s^2(T_C) = \mu_s^2 + c_s T_C^2 = - v^2(T_C) \sqrt{\lambda_h \lambda_s},
\end{equation}
and noting that from \refeq{eq:higgs_ewsb}, $v^2(T_C) = v^2_\text{EW} - \frac{c_h}{\lambda_h} T_C^2$, we have:
\begin{equation}
\mu_s^2 = -(v^2_\text{EW} - \frac{c_h}{\lambda_h}T_C^2)\sqrt{\lambda_h \lambda_s} - c_s T_C^2
\end{equation}
Since $c_s,\ c_h$ and $\lambda_h$ are expressed (cf.\ \refeqs{eq:c_h},\ref{eq:c_s} and \ref{eq:lambda_h}) in terms of $\lambda_m$, $\lambda_s$ and experimentally measured values, the region of the SSM under study is spanned by parameters $\{\lambda_s, \lambda_m, T_C\}$.

We constructed a benchmark point with a strong first-order phase
transition from a tree-level barrier, $\{T_C,\lambda_m,\lambda_s\} =
\{110\,\text{GeV}, 1.5, 0.65\}$. To find the nucleation temperature we
use\footnote{See e.g. Ref.\ \cite[Ch 4.4]{White:2016nbo}.},
\begin{equation}
  \frac{S_E(T_N)}{T_N} \approx 140,
  \label{Eq:Nucleation_temp}
\end{equation}
where $S_E(T)$ is the Euclidean action defined in
\refeq{eq:action}. Varying the temperature and using \bp to evaluate
the action, we found that this was reached at $T_N \approx 85$
GeV.

\begin{figure}[h!]
\centering
%  \begin{subfigure}[b]{0.49\textwidth}
    \includegraphics[width=0.49\textwidth]{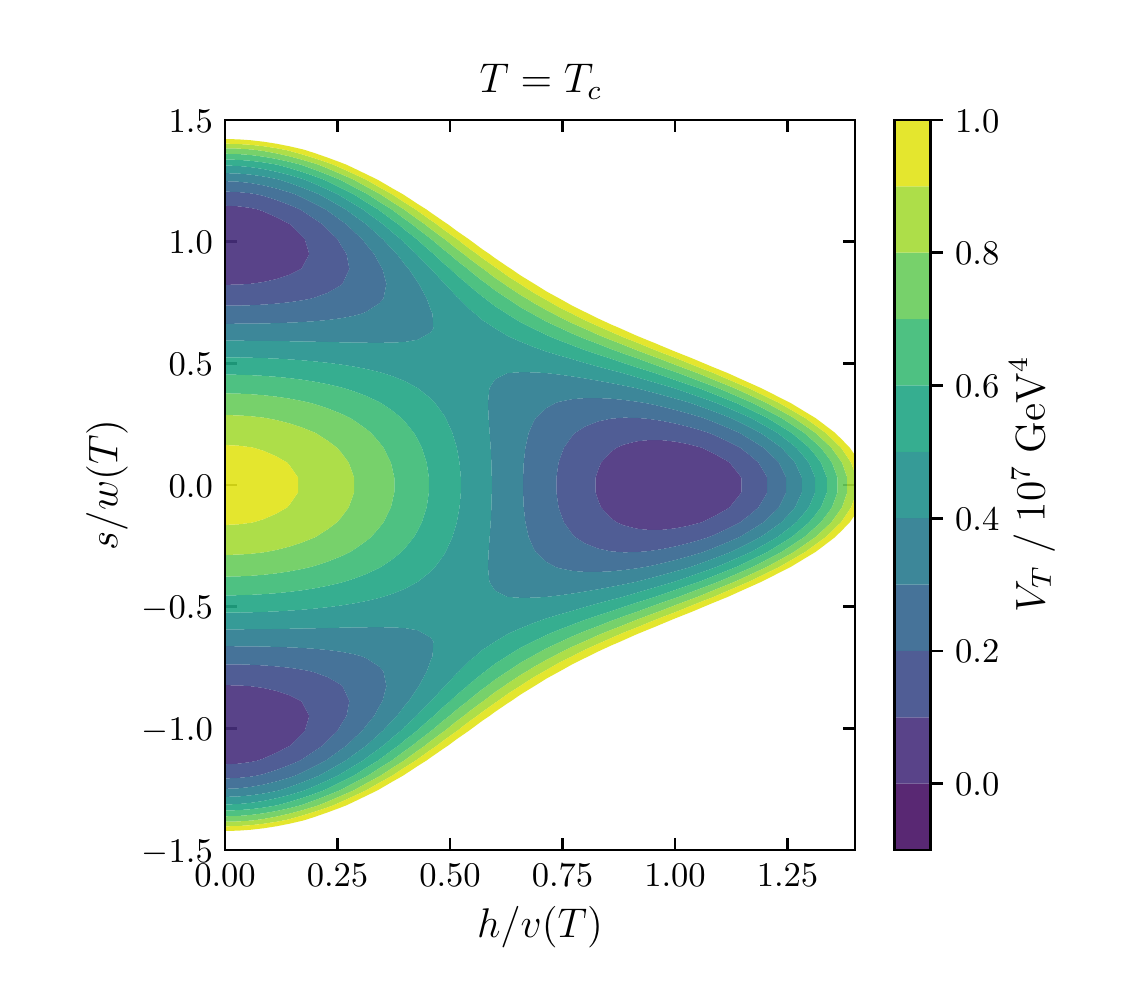}
    %\caption{$T=T_C$}
   %% \label{fig:fig_contour_tc}
%  \end{subfigure}
  %
%  \begin{subfigure}[b]{0.49\textwidth}
    \includegraphics[width=0.49\textwidth]{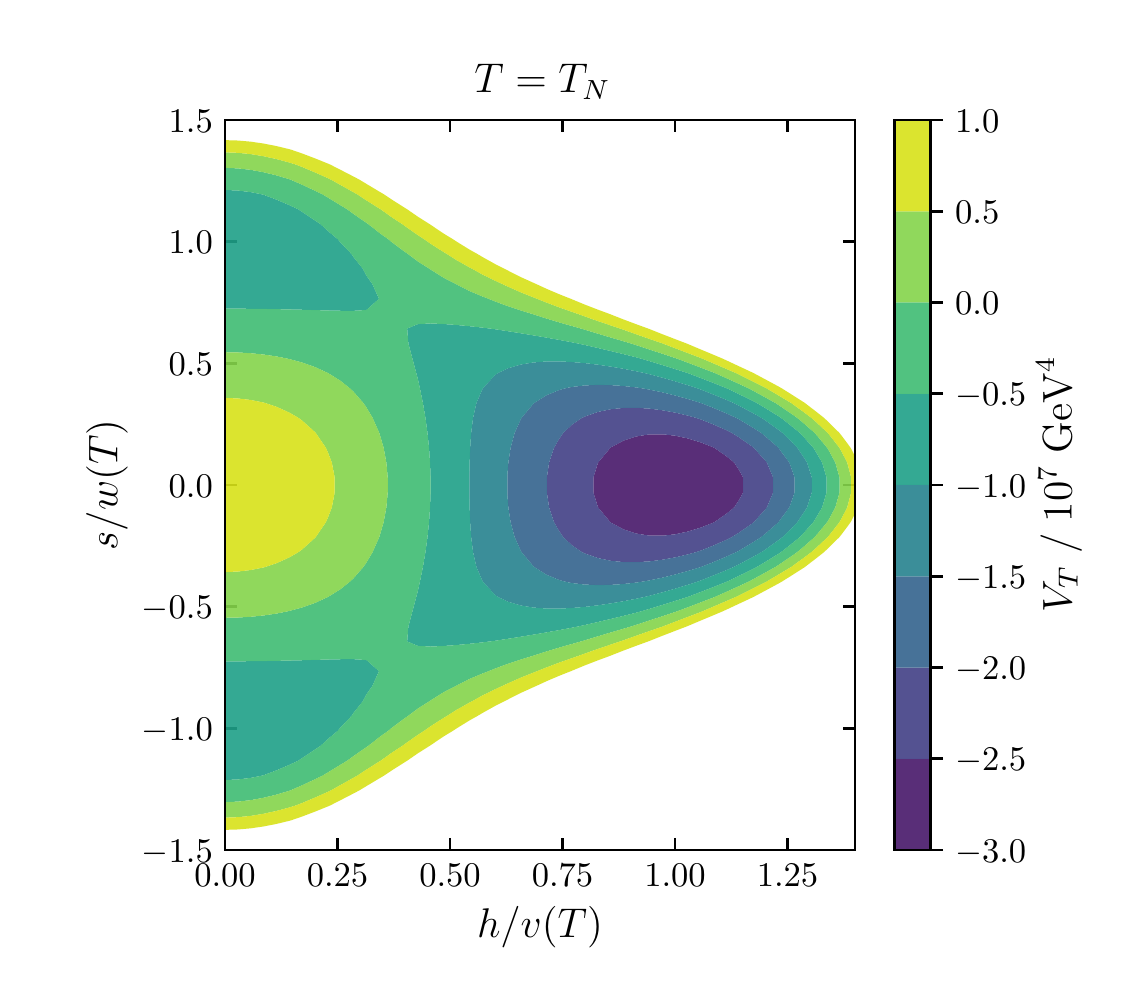}
    %\caption{$T=T_N$}
    %%\label{fig:fig_contour_tn}
%  \end{subfigure}
  \\
  %
%  \begin{subfigure}[b]{0.49\textwidth}
    \includegraphics[width=0.49\textwidth]{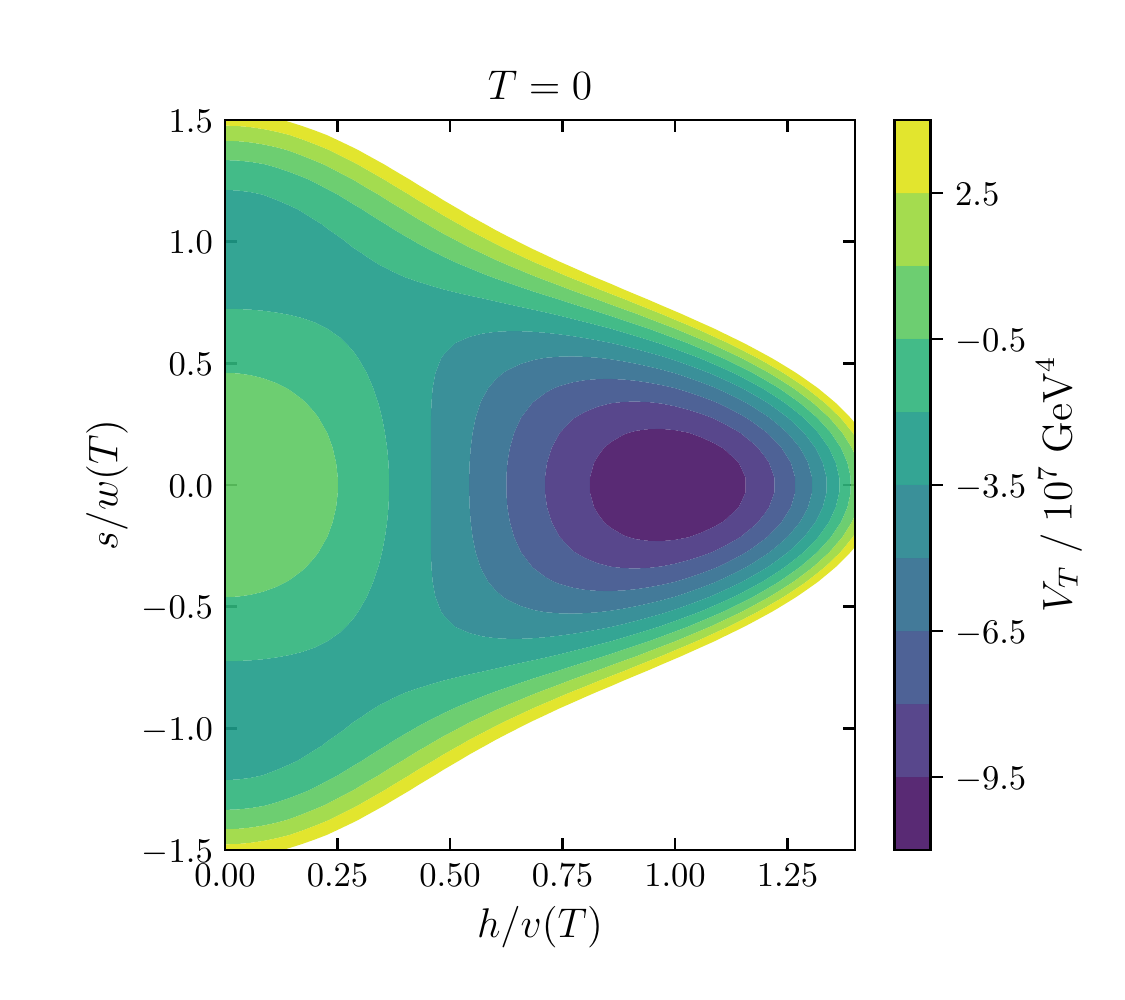}
    %%\caption{$T = 0$}
    %%\label{fig:fig_contour_0T}
%  \end{subfigure}
\caption{Contour plots of the effective potential for our benchmark SSM point, $\{T_c, \lambda_m, \lambda_s\} = \{110\ \textrm{GeV},1.5,0.65\}$, at the critical temperature $T = T_c$ (top left frame), at the nucleation temperature $T = T_N$ (top right frame) and at zero temperature $T = 0$ (bottom frame).}
\label{fig:sm_plus_singlet_contours}
\end{figure}

\figref{fig:sm_plus_singlet_contours} shows the structure of the
benchmark potential at $T = 0$, $T = T_N$, and $T = T_C$.  As
described earlier we are considering scenarios where there is a phase
transition between a minimum with non-zero $\langle s \rangle = w$ to
the electroweak symmetry breaking minimum with $\langle h \rangle = v$
and $\langle s \rangle = 0$.  In the top left frame of
\figref{fig:sm_plus_singlet_contours} we can see that the electroweak
minimum and the $\mathbb{Z}_2$ breaking minima\footnote{Note there are
  two of them due to the underlying symmetry.} are indeed degenerate at $T=T_c$,
which we have ensured by construction. Note that in this plot the
origin has already been destabilized, and is a local maximum.  As the
temperature cools the electroweak minimum becomes deeper and we reach
the nucleation temperature we have calculated using \bp and
\refeq{Eq:Nucleation_temp}, where the potential has the shape shown by
the color contour in the top right frame of
\figref{fig:sm_plus_singlet_contours}.  After the phase transition the
temperature continues to cool down and at $T=0$ the potential has the
shape shown in the bottom frame of \figref{fig:sm_plus_singlet_contours}
with an EWSB minimum of $v_\text{EW} = 246.22$ GeV.

\begin{figure}[h!]
\centering
  \begin{subfigure}[b]{0.49\textwidth}
    \includegraphics[width=\textwidth]{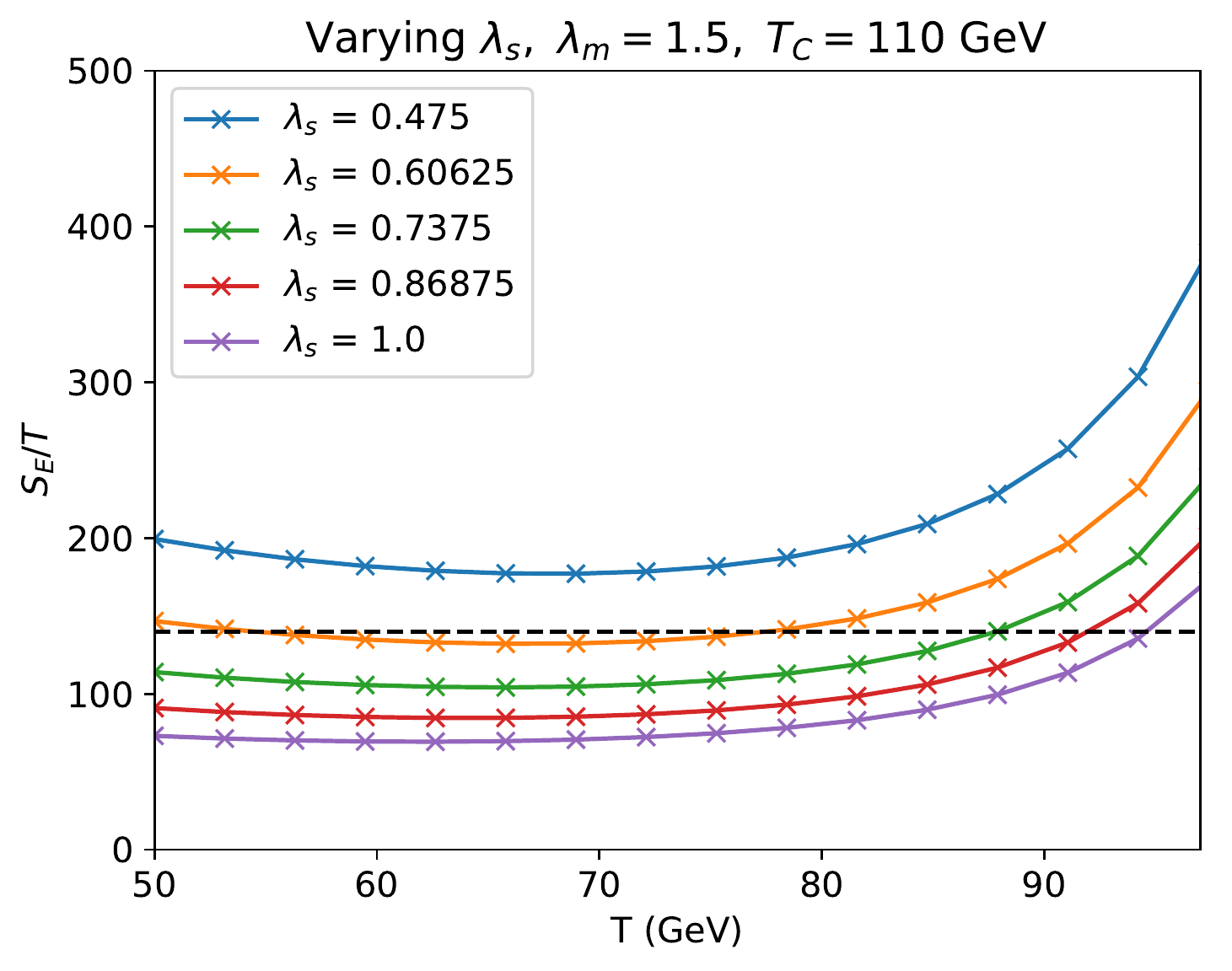}
    %%\label{fig:nucleation_lambda_s}
  \end{subfigure}
  \begin{subfigure}[b]{0.49\textwidth}
    \includegraphics[width=\textwidth]{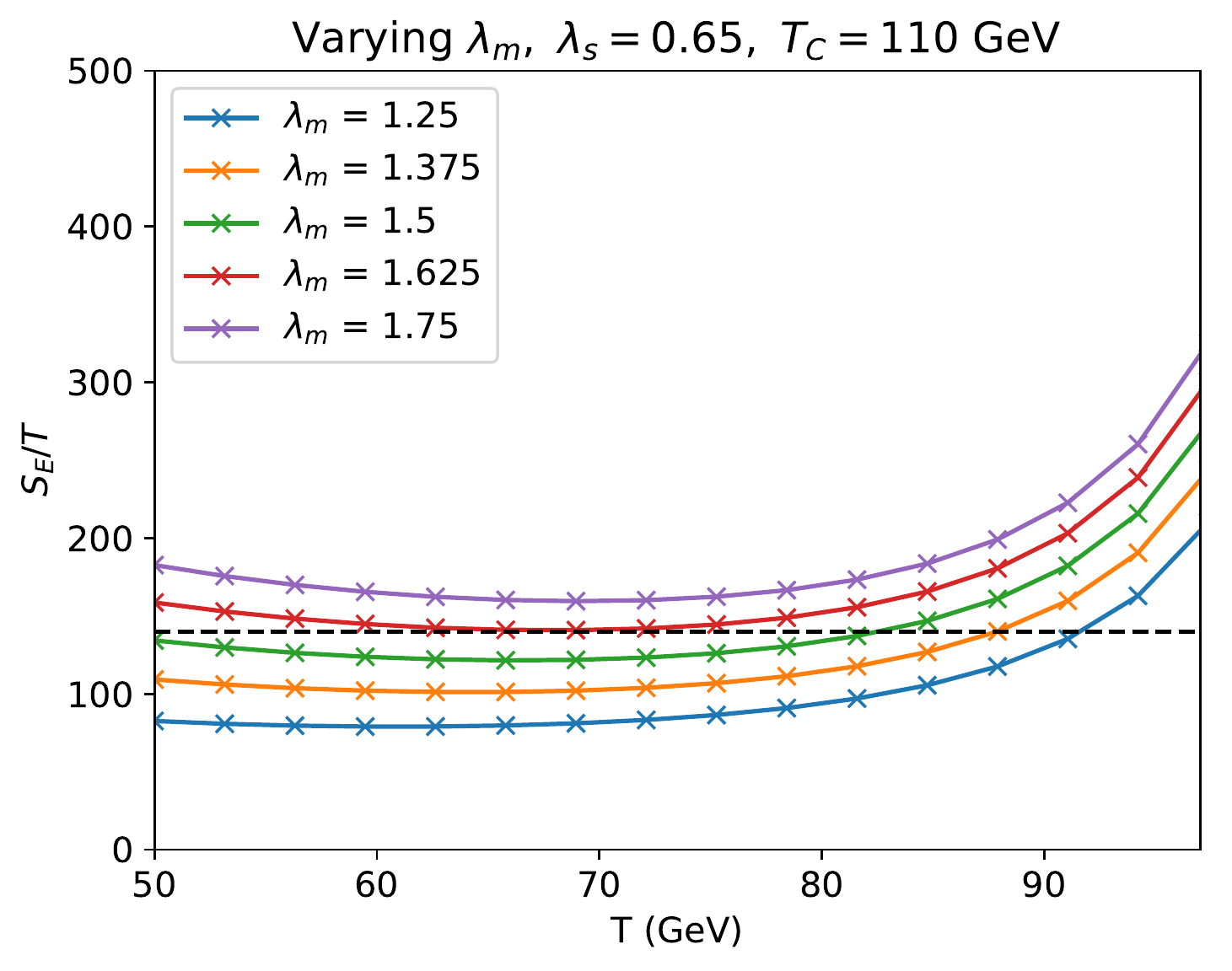}
    %%\label{fig:nucleation_lambda_m}
  \end{subfigure}
  \\
  \begin{subfigure}[b]{0.49\textwidth}
    \includegraphics[width=\textwidth]{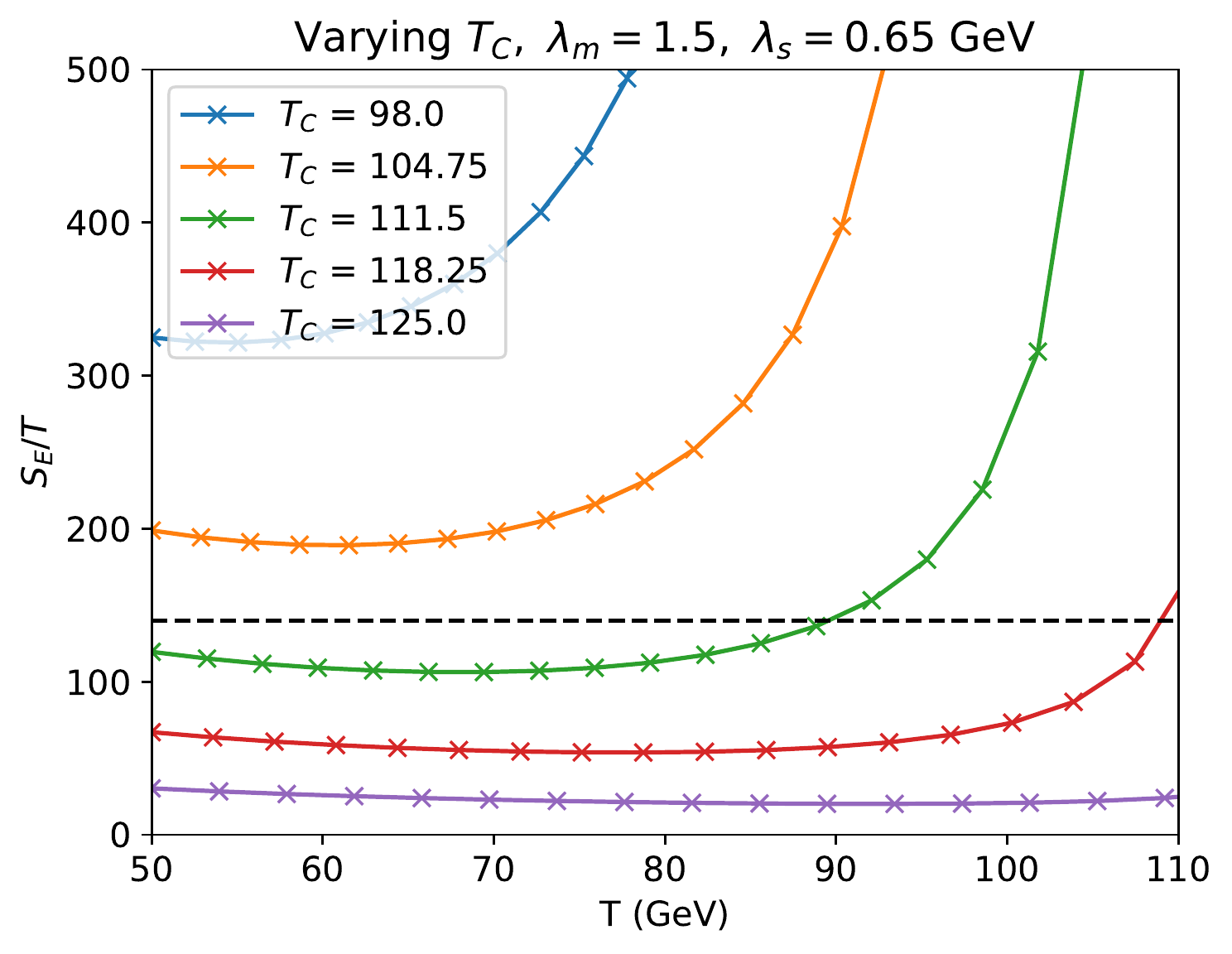}
    %%\label{fig:nucleation_Tc}
  \end{subfigure}
\caption{Change in Euclidean action as a function of temperature for points near our benchmark. We vary $\lambda_s$ (top left frame), $\lambda_m$ (top right frame) and the critical temperature, $T_C$ (bottom frame), relative to the base point $\{T_C,\lambda_m,\lambda_s\} = \{110\ \textrm{GeV},1.5,0.65\}$. The horizontal black dashed line represents the nucleation threshold $S_E / T = 140$, and the intersection of a curve with this line is the nucleation temperature. For all choices of the parameters as $T$ approaches $T_C$, the solution becomes thin walled and $S_E$ approaches infinity which causes the nucleation curves to diverge, though some curves are cut off before this is visible, and will run into numerical problems associated with thin walls for temperatures close enough to $T_c$. }
\label{fig:nucleation_plots}
\end{figure}

Having constructed the benchmark point, we now use \bp to investigate
the bounce action by individually varying the parameters
$\{T_C,\lambda_m,\lambda_s\}$. For each parameter set, \bp was used to
solve the bounce equation for temperatures in the interval
$[50\,\textrm{GeV}, T_C]$. The change in bounce action with respect to
temperature is shown in \figref{fig:nucleation_plots}.  We mark
$S_E(T)/T \approx 140$ by a horizontal line, showing how the
nucleation temperature changes with respect to each parameter in the
vicinity of the benchmark point.  As was first demonstrated in
\refcite{KurupDynamicselectroweakphase2017} a very significant
constraint on successful electroweak baryogenesis in this model is
whether or not the bubble nucleation takes place at any finite
temperature.  While we do not perform a full scan, by varying about
our benchmark we also find this constraint has an impact.

As can be seen in the top left frame \figref{fig:nucleation_plots} if
$\lambda_s$ is much smaller than that of the benchmark point we find
no solution to \refeq{Eq:Nucleation_temp}.  While our results are not
directly comparable to those of
\refcite{KurupDynamicselectroweakphase2017} due to different
approximations made in the calculation of the potential and slightly
different criteria for bubble
nucleation\footnote{\refcite{KurupDynamicselectroweakphase2017}
  require 100 instead 140 on the right hand side of
  \refeq{Eq:Nucleation_temp}.} these findings are in qualitative
agreement with their results, which are a result of smaller
$\lambda_s$ leading to a larger height and width of the barrier in the
first order phase transition. At the same time as shown in the top
right frame of \figref{fig:nucleation_plots} if $\lambda_m$ is too
large then there is also no bubble nucleation, and this result is
again in qualitative agreement with the findings of
\refcite{KurupDynamicselectroweakphase2017}.  Finally we see in the
bottom frame of \figref{fig:nucleation_plots} that for our benchmark
values of $\lambda_s$ and $\lambda_m$ if the critical temperature is
too large then there is again no bubble nucleation.  The scripts used
to run these tests are distributed under the
\code{examples/sm-plus-singlet} directory.

\section{Conclusions}

Vacuum decay appears in a wide variety of contexts in particle physics
and cosmology. For example there may be fundamental symmetries, like
the electroweak symmetry, that get broken in the cosmological history,
and may create observable gravitational waves or generate the observed
baryon asymmetry of the universe through an electroweak baryogenesis
mechanism. There is also the possibility of SM extensions with deeper
underlying charge or color breaking minima at zero temperature, where
the possibility of the vacuum decay can place limits on the model
depending on the lifetime of the metastable electroweak vacuum. 
Even the SM electroweak vacuum may be metastable if it is valid up to the
Planck scale.

In cases where the vacuum decays by bubble nucleation, calculating the rate of vacuum decay and related properties of the transition requires solving the bounce equations to obtain the bubble profile and Euclidean action.
We have described how this can
be done using \bp, a new C++ code, which is easy to use, fast and
adaptable. \bp is designed with flexibility and modularity in mind and is distributed
with two methods for solving the bounce equations: a perturbative method
capable of finding the bounce solution for any number of scalar
fields, and a special shooting method for one-dimensional potentials.
Each component in the perturbative calculation can be replaced and
updated, allowing both short term adaptations and long term evolution.

We tested \bp against existing codes, \cosmo and \ab, and against a
number of example potentials with known analytic solutions. We found
that \bp is fast and can find the bounce solution for a $3$ field
potential in under one second. For single field potentials, \bp is the fastest of the three codes. Finally, this new code is intended to grow and
develop. We strongly encourage users to give us feedback on the code, suggestions for new features, and to contact us with any questions relating to \bp.

\label{sec:conclusions}

\section*{Acknowledgments}

We thank Werner Porod, Ben O'Leary and Jose Eliel Camargo-Molina for
helpful discussions. We are also very grateful to Sujeet Akula for
his contributions in the early stages of this work.
The work of P.A., C.B., A.F.\ and D.H.\ was supported by the Australian
Research Council through the ARC Centre of Excellence for Particle Physics
at the Terascale (CoEPP) (grant CE110001104).  The work of P.A.\ is also
supported by the Australian Research Council Future Fellowship grant
FT160100274.  The work of D.H.\ was supported by the University of Adelaide,
and through an Australian Government Research Training Program Scholarship.
D.H.\ also acknowledges financial support from the Grant Agency of the Czech
Republic (GACR), contract 17-04902S, and from the
Charles University Research Center (grant UNCE/SCI/013).
TRIUMF receives federal funding via a contribution agreement with the National Research Council of Canada.

\appendix

\section{Interface to bubbler solvers --- \code{bubbler}}\label{sec:bubbler}

For ease of comparing results from \bp, \cosmo and \ab, we make our testing
suite publicly available at \bubblergitrepo.
This is a Python interface to \ab, \bp and \cosmo for solving the bounce action
and plotting the profiles.
This requires you to set paths to the codes
\begin{lstlisting}[language=bash]
export PYTHONPATH=Absolute/Path/To/CosmoTransitions
export BUBBLEPROFILER=Absolute/Path/To/BubbleProfiler
export ANYBUBBLE=Absolute/Path/To/AnyBubble
\end{lstlisting}
If you only wish to use particular codes, as e.g., you have not installed one, add the keyword argument for \code{backend}, e.g., \code{backends=["bubbleprofiler"]}.

To calculate the bounce actions, e.g.,
\begin{lstlisting}[language=python]
>>> from bubbler import bubblers
>>> print bubblers("0.1*((-x + 2)^4 - 14*(-x + 2)^2 + 24*(-x + 2))")
=== cosmotransitions ===
action = 52.5648686747
time = 0.272551059723
command = fullTunneling

=== anybubble ===
action = 52.3890699633
time = 6.020316
command = math -script /tmp/tmpp2sNls/anybubble.ws

=== bubbleprofiler ===
action = 54.112
time = 0.0439109802246
command = ~/BubbleProfiler/bin/run_cmd_line_potential.x --force-output --write-profiles --potential '0.1*((-x + 2)^4 - 14*(-x + 2)^2 + 24*(-x + 2))' --field 'x' --output-file /tmp/tmputKUDO --initial-step-size 0.01 --domain-start -1.0 --domain-end -1.0 --local-minimum 0.0 --global-minimum 5.0 --rtol-action 0.001 --rtol-fields 0.001 --integration-method runge-kutta-4 --n-dims 3  > /dev/null 2>&1
\end{lstlisting}
\code{bubblers} itself returns a dictionary-like object of information about the solutions. You can select $d=4$ by the keyword argument \code{dim = 4}. The potential may be a multi-field one.
For the profiles, the code
\begin{lstlisting}[language=python]
>>> from bubbler import profiles
>>> profiles("0.1*((-x + 2)^4 - 14*(-x + 2)^2 + 24*(-x + 2))")
\end{lstlisting}
shows the bubble profile for every field for every code.

\section{Multi-field polynomial potentials}\label{app:test_potentials}

We devised polynomial potentials for the purposes of testing with multiple
fields. For one field, we use the potential,
\begin{equation}
V(x) = \tfrac{1}{10}\left(x^4 -8 x^3 + 10 x^2 + 8\right).
\end{equation}
with extrema at $x = 0$, $1$ and $5$. For greater than one field, we construct potentials of the form
\begin{equation}
V = \left( \left[\sum_{i=1}^n c_i (x_i - 1)^2\right] - c_{n+1} \right)\left(\sum_{i=1}^n x_i^2\right),
\end{equation}
with vacua close to $x_i = 0$ and $1$ for all $i$. Thus to construct potentials
with $2$--$8$ fields, we pick the coefficients,
\begin{align}
c&=\left(1.8, 0.2, 0.3\right), \label{eq:2_field_potential}\\
c&=\left(0.684373, 0.181928, 0.295089, 0.284821\right),\\
c&=\left(0.534808, 0.77023, 0.838912, 0.00517238, 0.258889\right),\\
c&=\left(0.4747, 0.234808, 0.57023, 0.138912, 0.517238, 0.658889\right),\\
c&=\left(0.34234, 0.4747, 0.234808, 0.57023, 0.138912, 0.517238, 0.658889
\right),\\
c&=\left(0.5233, 0.34234, 0.4747, 0.234808, 0.57023, 0.138912, 0.517238,
0.658889\right),\\
c&=\left(0.2434, 0.5233, 0.34234, 0.4747, 0.234808, 0.57023, 0.138912,
0.51723, 0.658889\right).
\end{align}
This form was introduced for two-dimensional potentials in
\refcite{Wainwright:2011kj}. The parameter $c_{n+1}$ governs the degeneracy
of the vacua; for $c_{n+1} \ll 1$, the true and false vacua are almost
degenerate and the bubble profile must be thin-walled.

\section{Command line interface}
\label{sec:User-Options}
The main user interface for \bp is the \code{bin/run_cmd_line_potential.x} tool.
An example command demonstrating the minimal required inputs is:

\begin{lstlisting}[language={}]
run_cmd_line_potential.x --potential "(x^2 + y^2)*(1.8*(x - 1)^2 + 0.2*(y - 1)^2 - 0.3)" --field "x" --field "y" --false-vacuum-at-origin --global-minimum 1.0402967171 1.53520719837
\end{lstlisting}
which results in output
\begin{lstlisting}[language={}]
Potential: (x^2 + y^2)*(1.8*(x - 1)^2 + 0.2*(y - 1)^2 - 0.3)  
Field: x
Field: y
# Action: 20.8363
\end{lstlisting}

A comprehensive set of options allows the user to choose which algorithms will
be used to solve the bounce equations, customize relevant parameters, and
specify output formats. We list these below in three groups: general options
applying to all potentials, options which affect the ansatz solution, and
options specific to solving single field problems with the direct shooting
algorithm.

\subsection{General options}

\begin{description}
\item[\texttt{-\/-help}] Print a summary of the command line options.

\item[\texttt{-\/-potential} (required):] Potential for one or more fields given
  in GiNaC's \cite{BauerIntroductionGiNaCFramework2002} syntax. For example,
  the two field potential in equation \refeq{eq:2_field_potential} would be
  specified by:

\begin{lstlisting}
--potential "(x^2 + y^2)*(1.8*(x - 1)^2 + 0.2*(y - 1)^2 - 0.3)"
\end{lstlisting}

\item[\texttt{-\/-field} (required):] Indicates which symbols in the potential
  correspond to fields. Can be specified multiple times. For example, a
  two field potential with fields \code{x} and \code{y} would require:

\begin{lstlisting}
--field "x" --field "y"
\end{lstlisting}

\item[\texttt{-\/-n-dims}:] Number of spacetime dimensions $d$. Typically
  $d = 4$ for zero temperature calculations, and $d = 3$ at finite
  temperatures. Corresponds to the parameter $n$ in
  \secref{sec:PhysicalProblem} via $n = d - 1$.

\item[\texttt{-\/-perturbative}:] Force \bp to always use the perturbative
  algorithm described in \secref{sec:perturbative_algorithm}. If this option
  is not specified, the direct shooting method described in
  \secref{sec:shooting} will be used for single field potentials.

\item[\texttt{-\/-global-minimum} (required): ] Location of the true vacuum.
  The potential in \refeq{eq:2_field_potential} has a true vacuum at
  \code{x=1.0402967171, y=1.0402967171}, which would be specified by:

\begin{lstlisting}
--global-minimum 1.0402967171 1.0402967171
\end{lstlisting}

The order of the field coordinates should be the same as the \code{--field}
flags. If a true vacuum is not specified, \bp will attempt to find it using
global optimization.

\item[\texttt{-\/-opt-timeout}:] Sets a time limit if finding the true vacuum
  using global optimization. Omitting this option, or specifying a value of
  $0$ results in no time limit.

\item[\texttt{-\/-local-minimum} (required):] Location of the false vacuum.
  Specification format is the same as for \code{--global-minimum}. Required
  unless the \texttt{-\/-false-vacuum-at-origin} flag is given.

\item[\texttt{-\/-false-vacuum-at-origin}:] Assume that the false vacuum lies
  at the origin in field space.

\item[\texttt{-\/-domain-start}:] Radial coordinate for start of finite domain
  ($\rhomin$) on which bounce equations are solved. Omitting this option, or
  specifying a negative number will cause \bp to guess an
  appropriate value. This is done by finding the point closest to the origin
  where the radial derivative of the ansatz solution is equal to $10^{-5}$.

\item[\texttt{-\/-domain-end}:] Radial coordinate for end of finite domain
  ($\rhomax$). Omitting \texttt{-\/-domain-end} or specifying a negative value
  will cause \bp to estimate a value by finding the outermost point
  at which the value of the ansatz solution is less than or equal to $10^{-5}$.
  This is usually sufficient, but in the case of 'long tailed' solutions
  automatic domain sizing may cause the action to be underestimated.  We
  recommend that users relying on automatic domain sizing for large scans
  take care to verify that manually increasing the domain size does not
  significantly change the calculated action.

\item[\texttt{-\/-initial-step-size}:] This option specifies the approximate initial step size to use in solving ODEs. When using the perturbative algorithm, a fixed step size close to this value will be used over the radial grid. The shooting algorithm uses an adaptive step size which will evolve from this initial value during integration.

\item[\texttt{-\/-interpolation-fraction}:] Approximate fraction of the total
  number of grid points to use when building cubic spline interpolations of
  intermediate solutions.

\item[\texttt{-\/-integration-method}:] Algorithm to use when integrating the
  perturbation equations, \refeq{eq:perturbations_matrix}. Available options are
  fourth order Runge-Kutta (\code{--integration-method=RK4}) or the Euler
  method (\code{--integration-method=euler}).

\item[\texttt{-\/-rtol-action}, \texttt{-\/-rtol-field}:] Relative tolerance
  criteria for determining when to halt iteration. \code{--rtol-action} is
  compared to the relative change in the bounce action between iterates.
  \code{--rtol-fields} is compared to the relative change in the starting
  values $\phi_{0 j} = \phi_j(\rhomin)$ of each field. The iteration halts when
  both of these quantities fall below their respective thresholds.

\item[\texttt{-\/-max-iterations}:] Maximum number of iterations to perform,
  regardless of convergence criteria. Omitting this option or setting it to a
  negative value will cause \bp to iterate until the convergence
  criteria set via \code{--rtol-action} and \code{--rtol-field} are met.

\item[\texttt{-\/-output-path}:] Directory in which to store output files.
  Three files are created:
\begin{itemize}
\item \code{action.txt} File listing the Euclidean action of the solution at
  each iteration, up to the final profile.
\item \code{field_profiles.txt} Listing of the field profiles for each
  iteration.
\item \code{perturbations.txt} Listing of the correction functions applied at
  each iteration.
\end{itemize}

\item[\texttt{-\/-force-output}:] Overwrite files in the output directory. If
  this option is not specified and files are present, \bp will exit with an
  error message.

\item[\texttt{-\/-output-file}:] Write an additional summary file at the
  indicated location. By default, this contains only the Euclidean action of
  the final solution.

\item[\texttt{-\/-write-profiles}:] Print the final set of field profiles to
  the console after execution. If the \code{--output-file} option is
  specified, the profiles will be written to the summary file instead.

\item[\texttt{-\/-verbose}] Print detailed information to the console while
  computing the bubble profile.

\end{description}

\subsection{Ansatz options} \label{sec:ansatz_cli_options}

By default, the parametric kink ansatz described in \secref{sec:ansatz} is used to construct the initial field profiles. The following options provide alternative ansatz solutions.

\begin{description}

\item[\texttt{-\/-shooting-ansatz}:] Use the 1D direct shooting method
  described in \secref{sec:shooting} to construct the initial profiles. Note
  that the options described in \secref{sec:1D_shooting_options} will affect
  the calculation of this ansatz.

\item[\texttt{-\/-ansatz-file}:] Load a precomputed ansatz from a text file.
  Each row consists of a radial field coordinate $\rho_j$, followed by the
  field values $\phi_i(\rho_j)$. Columns are separated by spaces.  The order
  in which the fields appear must match the \code{--field} specifications,
  and the radial coordinates $\rho_j$ must match the discrete grid
  representation to a tolerance of $10^{-6}$, if a grid has already been
  given.  An example using a textfile ansatz is provided in the
  \code{examples/textfile-ansatz} directory.

\end{description}

\subsection{Options specific to single field direct shooting}
\label{sec:1D_shooting_options}

The following options will only have an effect on the behavior of \bp when
solving single field problems using the direct shooting method.

\begin{description}

\item[\texttt{-\/-barrier}:] Specify the location of the barrier separating
  the true and false vacua. If not supplied, \bp will attempt to find the
  barrier numerically.

\item[\texttt{-\/-action-arrived-rel}:] Relative tolerance for arriving at the
  false vacuum when calculating bounce action.

\item[\texttt{-\/-shoot-ode-abs}, \texttt{-\/-shoot-ode-rel}:] Absolute and
  relative error tolerances for ODE integrator when calculating bubble profile.

\item[\texttt{-\/-action-ode-abs}, \texttt{-\/-action-ode-rel}:] Absolute and
  relative error tolerances for ODE integrator when calculating bounce action.

\item[\texttt{-\/-drho-frac}:] Initial step size relative to characteristic
  bubble scale (\refeq{eq:bubble_scale}).

\item[\texttt{-\/-evolve-change-rel}:] How far to evolve using approximate
  analytic solutions (\refeq{eq:approx_gt} and \refeq{eq:approx_lt}) in terms
  of change in field value relative to difference between true and false
  vacua. Corresponds to the $f$ parameter in \refeq{eq:critical}.

\item[\texttt{-\/-bisect-lambda-max}:] Maximum value of bisection parameter
  $\lambda$ (\refeq{eq:thin_wall_reparam}).

\item[\texttt{-\/-iter-max}:] Maximum number of iterations of the shooting method.

\item[\texttt{-\/-periods-max}:] Evolve the field for a maximum of this number
  of multiples of the characteristic bubble scale (\refeq{eq:bubble_scale}).

\item[\texttt{-\/-f-y-max}:] As discussed in \secref{sec:evolve}, we evolve the
  field analytically until it changes by a specified fraction. This involves
  evaluating complicated functions, such as inverse sinc and hyperbolic sinc,
  at $t$ (defined in \refeq{eq:measure}). If $t > \code{f_y_max}$, we instead
  use an asymptotic approximation for the critical time and velocity at that
  time.

\item[\texttt{-\/-f-y-min}:] Similar to the \code{--f-y-max} argument: if
  $t < \code{f_y_min}$, we use an asymptotic approximation for the velocity at
  the critical time.

\item[\texttt{-\/-y-max}:]  Similar to the \code{--f-y-max} argument: we find
  the action for the interior of the bubble, before the field changes by a
  specified fraction, analytically. If $m \hat \rho > \code{y_max}$, we use a
  simpler  asymptotic approximation for it.

\end{description}

\clearpage

%\section*{References}
\addcontentsline{toc}{section}{References}

\bibliographystyle{JHEP}
\bibliography{BubbleProfiler}

\end{document}